\documentclass{cambridge6A}
    \usepackage{graphicx}
\usepackage{color}
             \usepackage{amsmath}
           \usepackage{amsfonts}
           \usepackage{amssymb}

\begingroup\makeatletter\ifx\SetFigFont\undefined%
\gdef\SetFigFont#1#2#3#4#5{%
  \reset@font\fontsize{#1}{#2pt}%
  \fontfamily{#3}\fontseries{#4}\fontshape{#5}%
  \selectfont}%
\fi\endgroup%
 
\begin{document}

\author[Donald Marolf]{Donald Marolf}

\chapter{Black holes and branes in supergravity \\ }

\contributor{Donald Marolf \affiliation{Department of Physics, UCSB, Santa Barbara, CA 93106, USA}}

\section{Introduction}
\label{intro}

The goal of the present chapter is to introduce black holes and branes in supergravity in the simplest possible manner.  As a result, we make no attempt to be complete and, in fact, we will purposely leave out many points dear to the hearts of practicing string theorists and super-gravity experts.  In particular, spinors and supersymmetry will make only a passing appearance in section \ref{bps}, which can be skipped if the material seems too technical (though be sure to read the Executive Summary at the end of that section).   Instead, we  focus on bosonic spacetime solutions and the dynamics of the associated bosonic fields.

Even within this limited scope, our referencing of the original works
will be rather spotty.  The interested reader can consult \cite{YR} for a more encyclopedic review of branes and black holes in string theory (as of 1997) with
numerous references  to the original works.  We also refer the reader to \cite{45rev} for a more recent review of black holes in 4- and 5-dimensional supergravity and for complimentary material in 10- and 11-dimensions, to \cite{ResLet} for a partial guide to the literature as of 2004, and to various textbooks \cite{GSW,Joe,CVJ,BBS,DineBook,Kbook} for further reading.

Our treatment will focus on supergravity theories in 10- and 11-dimensions which are in many ways simpler than their lower-dimensional counterparts and which allow us to make direct contact with string theory.
We draw heavily on Polchinski's treatment \cite{Joe}, though the style is (hopefully)
more adapted to the current audience.   We begin in
section \ref{11sugrav} with a discussion of both the kinematics and dynamics of 11-dimensional
supergravity  and then introduce the
associated branes in section \ref{Mbr}.  We follow tradition in referring to 11-dimensional supergravity as ``M-theory\footnote{The origin of this term nevertheless seems to be lost in the mists of history.}."

The case of ten dimensional supergravity is also central to our
mission.  We choose to approach this subject via Kaluza-Klein reduction of
eleven dimensional supergravity.  Section \ref{KKred} thus begins with
some introductory remarks on Kaluza-Klein compactifications
and then constructs a supergravity theory (the type IIA theory)
in 9+1 dimensions.  Section \ref{10branes} introduces the branes
of type IIA theory and then discusses both the very similar case of type IIB supergravity and the so-called T-duality symmetry that relates the two
theories.  The type I and heterotic theories are mentioned only briefly.  We close with some brief remarks on D-brane
perturbation theory and a few words about black hole
entropy via D-branes in section \ref{pert}.  We hope this provides a useful basis for the later chapters in this volume.

\section{Supergravity in eleven dimensions}
\label{11sugrav}

Before diving into the details, a few words of orientation are
in order.  We will shortly see that supergravity in
eleven (10+1) dimensions is really not much more complicated than the 3+1
Einstein-Maxwell theory of Einstein-Hilbert gravity coupled to
Maxwell electrodynamics.   The same is not
as true of supergravity in lower dimensions.  In ten  (9+1)
dimensions and below, many interesting supergravity theories
contain a so-called dilaton field which couples non-minimally to
the Maxwell-like gauge fields.  As a result, the equivalence principle
does not hold in such theories and different fields couple to
distinct metrics that differ by a conformal factor.
However, in eleven dimensions, properties of the supersymmetry
algebra guarantee that any supergravity theory
containing no fields
with spin higher than two\footnote{Except
for anti-symmetric tensor fields, which propagate on curved manifolds
without the constraints associated with other higher spin fields.  Because of these constraints, theories with spin higher than two are generally believed to be inconsistent unless they include an infinite number of such fields.} has
no dilaton.  In fact, there is  a unique supergravity theory in
eleven dimensions and it contains only three fields:  the metric,
a U(1) (i.e., abelian, Maxwell-like) gauge field, and a spin 3/2
gravitino.

In contrast, in $D > 11$ spacetime dimensions there are {\it no} supergravity theories without fields of spin $s > 2$. The basic reason for this property is that supersymmetry is associated with generators having spin 1/2. The action of such `supercharges' on a given field thus returns a field of different spin.  Working through the details one finds that, in any dimension $D$, theories with only spins $s \le 2$ can accommodate no more than 32 supercharges. In asymptotically flat settings, asymptotic Lorentz symmetry implies that each supercharge is associated with some component of a spinor of the associated Lorentz group. It turns out that 11-dimensional Majorana spinors have precisely 32 components, so that precisely one $D=11$ spinor worth of supercharges is allowed.  Any supergravity theory with 32 supercharges is called maximally supersymmetric.

Another reason to begin in 11 dimensions is that lower dimensional supergravity theories can generally be obtained through the Kaluza-Klein mechanism
in which some subset of the dimensions are taken to be compact
and small (and by also using certain so-called ``dualities'').  This mechanism will be discussed in sections
\ref{KK} and \ref{SUGRAVKK} below.

\subsection{On $n$-form gauge fields}
\label{forms}

We first address just the bosonic part of eleven-dimensional supergravity,
setting the fermionic fields to zero.  The differences between
this truncated
theory and 3+1 Einstein-Maxwell theory amount to just differing numbers
of dimensions.  This happens in two ways:  The first is
the obvious fact that
the theory lives in a 10+1 spacetime instead of a 3+1 spacetime.  The
second is that the gauge field is itself slightly `larger' than that of Maxwell
theory.  Instead of having a {\it vector} (or, equivalently, a one-form)
potential, the potential is a 3-form: $A_3$.

We will encounter a number of $n$-form potentials below.  Although they
may at first seem unfamiliar, they are in fact a very natural
(and very slight) generalization of Maxwell fields.  An $n$-form
gauge potential $A_n$ is associated with an $(n+1)$-form
field strength of the form $F_{n+1} = dA_n$, where $d$ is the exterior
derivative.  As a result, the field strength satisfies a Bianchi
identity $d F_{n+1} =0$.   As with the familiar Maxwell field, there is
an associated set of gauge transformations
\begin{equation}
A_n \rightarrow A_n + d \Lambda_{n-1},
\end{equation}
where $\Lambda_{n-1}$ is an arbitrary $(n-1)$ form.   Such
gauge transformations leave the field strength $F_{n+1}$ invariant.  For reference purposes, we record our conventions for differential forms (which agree with \cite{Joe}):
\begin{equation}
A_n = \frac{1}{p!} A_{\alpha_1 ...\alpha_n} dx^{\alpha_1} \wedge ... \wedge
dx^{\alpha_n},
\end{equation}
so that we have
\begin{equation}
\int A_n = \int A_{0123...(n-1)} d^nx.
\end{equation}

In $D$ spacetime dimensions (generally taken to be $11$ in this section), the
equation of motion for such a gauge field is typically of the
form
\begin{equation}
\label{current}
d\star F_{D-(n+1)} = \star J_{D-n},
\end{equation}
where, in a slight abuse of notation,
$\star F_{D-(n+1)}$ denotes the $D-(n+1)$ form that is the Hodge-dual  of
$F_{(n+1)}$ defined by
\begin{equation}
\star F_{\alpha_1 \dots \alpha_{d-p}} = \frac{1}{p!} \epsilon_{\alpha_1 \dots \alpha_{d-p}}{}^{\beta_1 \dots \beta_p} F_{\beta_1 \dots \beta_p},
\end{equation}
and $\epsilon$ is the Levi-Civita tensor.  Similarly,
$\star J_{D-n}$ is the Hodge-dual of some $n$-form current $J_n$.  These conventions are consistent with taking
the natural coupling in an action between an $n$-form
gauge field and its current to be of the form $\int_{\cal M} A_n
\wedge \star J_{D-n} \propto \int_{\cal M}\sqrt{-g} A_{\alpha_1\dots\alpha_n}J^{\alpha_1\dots\alpha_n}$, where ${\cal M}$ denotes the spacetime manifold.

As usual, gauge symmetry implies that the current is
conserved.  However, current conservation for a $(D-n)$-form current with
$D-n>1$ is, in a certain sense, a much stronger statement than
conservation of the current in 3+1 Maxwell theory.
Note that the analogue of Gauss'
law in the present context is to define the charge $Q_B$ contained in
a $(D-n)$-ball $B$ by the integral
$Q_{D-n} = \int_{\partial B} \star F_{D-(n+1)}$ over the boundary $\partial B$
of that ball.  Now, suppose that the current $J_{D-n}$ in fact
vanishes in a neighborhood of the surface $\partial B$.  Then by stokes
theorem and equation (\ref{current}) we can deform the surface $\partial B$ in
any way we like and, as long as the surface does not encounter any
current, the total charge $Q_{D-n}$ does not change.

Now, in
familiar 3+1 Maxwell theory, electric charge is measured by integrals over
2-surfaces.  This is associated with the fact that
an electrically charged particle sweeps out a worldline in spacetime.
Note that any sphere which can be collapsed
to a point without encountering the worldline of the particle must
enclose zero net charge.  The important fact is that,
in four dimensions, there are two-spheres which `link' with
any curve and which cannot in fact be shrunk to a point without encountering
the particle's worldline.
In contrast, circles do not link with worldlines in
3+1 dimensions.  For this reason, particles in 3+1 dimensions cannot be electrically
charged under any gauge field whose field strength is, for example, a
3-form.   This illustrates a general relation between a gauge field
and the associated charges:
unless the world-volume of an object can link with
surfaces of dimension $D-(n+1)$, it cannot be electrically charged under
an $n$-form gauge potential $A_n$.

While we are here, we may as well work out the relevant counting in general.  Let us
suppose that we have an $n$-form gauge field $A_n$ in $D$ spacetime
dimensions.  Then, we must integrate $\star F_{D-(n+1)}$ over a $D-(n+1)$
surface in order to calculate the charge.  Now, in $D$ dimensions,
surfaces of dimensions $k$ and $m$ can link without intersecting if $k+m +1 =D$ (i.e.,
curves and curves in three dimensions, 2-surfaces and worldlines in
four dimensions, etc.).  Thus, non-zero electric charge of $A_n$ is
associated with $n$ dimensional worldvolumes.  Such objects are generically
known as `$p$-branes' (as higher dimensional generalizations
of the term membrane). \index{p-brane} Here, $p$ is a the number of spatial
dimensions of the object; i.e., the electric charge of
an $n$-form gauge potential is carried by $(n-1)$-branes, whose
world volume has $n-1$ spatial dimensions and time.  This is how
strings, membranes, and other branes will arise in our discussion
of supergravity.

Note that, although $p$-branes are extended objects, the concept
of a charge {\it density} of $A_{p+1}$- charge on a $p$-brane is not appropriate.  Recall
that the charge is measured by any $D-(p+2)$ surface surrounding the
brane and that, by the above charge conservation argument, smooth deformations of the surface do not change the charge so measured.
Thus, the equations of motion tell us that moving the $D-(p+2)$
surface {\it along} the brane cannot {\it ever} change the flux through
the surface;  see Fig. 11.1 below.
Thus, non-uniform `pure' $p$-branes cannot exist!
The proper concept here is to assign to such a $p$-brane only
one number, the total charge.  It simply happens that the particular
type of charge being measured is somewhat less local than
the familiar electric charge; it is fundamentally associated with
$p+1$ dimensional hypersurfaces in the spacetime.\footnote{Of course, it is possible for a $p$-brane to also carry $A_n$-form charges with $n < p+1$.  E.g., a string can carry point-charges in addition to some intrinsic string charge.  In this case the proper concept is of a density of $A_n$-charge per unit $(p-n-1)$-dimensional volume.  Such charge densities need not be homogeneous. }

\vbox{
\centerline{\includegraphics[width=3in]{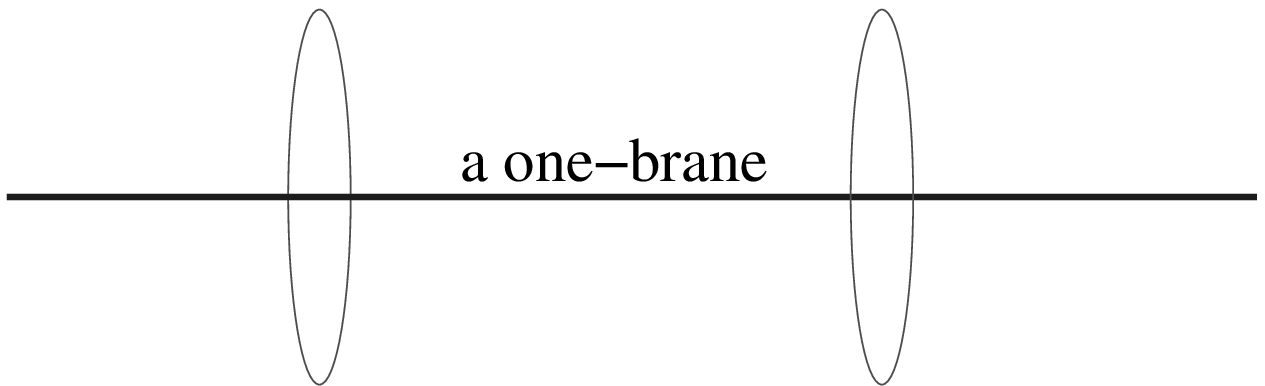}}
\centerline{Figure 11.1 By charge conservation, both circles necessarily
capture the same flux.}}

\medskip

As a small complication, we will be interested not only in electric
charges, but also in magnetic charges.  Indeed, in supersymmetric
string theory, both electric and magnetic charges appear to be on
an equal footing.  A useful point of contact for the present discussion is
to realize that, in a certain sense, both electrically and magnetically
charged `objects' occur in pure Einstein-Maxwell theory without any
matter fields.  These are just the electrically and magnetically
charged eternal black hole solutions.  Although the Maxwell field
satisfies both $dF=0$ and $d\star F=0$ at every non-singular point of
such spacetimes, the black holes can still be said to `carry charge' due to
topological effects:  the electric or magnetic flux starts in one
asymptotic region, funnels through the Einstein-Rosen bridge at
the `throat' of the black hole, and out into the other asymptotic
region\footnote{As a result, the electric charge of the black hole measured in
one asymptotic region is the opposite of the charge measured in the
other asymptotic region.  However, this need not trouble us so long
as we understand that we must first orient ourselves by
picking an asymptotic region in order to discuss  the notion of charge.}.
Note that black holes (i.e., point-like or zero-brane objects) may carry
both electric {\it and} magnetic charge for the Maxwell field in 3+1 dimensions.

The counting of dimensions
for magnetic charges proceeds much like the counting for electric
charges.  To define what we mean by a magnetic charge, we recall
that hodge duality $F \rightarrow \star F$ in Maxwell theory interchanges
electric and magnetic charge.  Thus, since electric charge is
associated with integrals of $\star F_{D-(n+1)}$, magnetic charge is defined
by integrating the field strength $F_{n+1}$ itself over an $n+1$
surface.  In $D$ dimensions, an $n+1$ surface can link with
$D-n -2$ worldvolumes, or $(D -n -3)$-branes.  As a check, for
3+1 Maxwell theory, we have magnetic $4- 1 - 3=0$-branes.

Let's take a look at the eleven-dimensional context.  Without knowing
anything more about supergravity than we already do, we can expect
two types of `objects' to be of particular interest from the point of
view of the 3-form gauge field $A_3$.  There may be 2+1 electrically
charged objects (2-branes) and $(D-n-3) +1 = (11-3-3)+1 = 5+1$ dimensional
magnetically charged objects (5-branes).  Since there are no
explicit charges in the theory, these `objects' (if they
exist) must be black-hole like `solitonic' solutions.  We will see
below that black two-brane and black five-brane solutions carrying
the proper charges do indeed exist in eleven-dimensional supergravity.
What is more, and what is different from lower dimensional
supergravity, is that the horizons of these black branes
remain smooth in the extreme limit
of maximal electric or magnetic charge.  The extremal versions of these
brane-solutions are what are usually referred to as `the M-theory
two-brane' or `M2-brane,'  and `the M-theory five-brane' or
`M5-brane.'  We will discuss these
in more detail in section \ref{branes}.

\subsection{Dynamics}

I hope the discussion of section \ref{forms} has provided
some orientation to supergravity in eleven dimensions.  Now,
however, it is time to fill in a few details.  For example,
it is appropriate to write down the full dynamics of the
system.  This is conveniently summarized by the action \cite{CJS}

\begin{eqnarray}
\label{eleven}
S&=& \frac{1} {2 \kappa_{11}^2 } \int d^{11} x \  e \Bigl[
\bigl( R - \frac{1}{2} |F_4|^2 \bigr)  \cr
&-& \frac{1}{2^3 \cdot 4!} (\overline{\psi}_\alpha\Gamma^{\alpha \beta
\gamma \delta \sigma \lambda}
\psi_\lambda + 12 \overline{\psi}^\beta \Gamma^{\gamma \delta} \psi^\sigma)
(F + \hat F)_{\beta \gamma \delta \sigma} \cr
&-& \overline{\psi}_\alpha
\Gamma^{\alpha \beta \gamma}
D_\beta\bigl(\frac{1}{2}(\omega + \hat \omega) \bigr)\psi_\gamma \Bigr]
- \frac{1}{12\kappa_{11}^2}
\int A_3 \wedge F_4 \wedge F_4,
\end{eqnarray}
where
\begin{equation}
\label{Fnorm}
|F_{m}|^2 = \frac{1}{m!}  g^{{\alpha_1} {\beta_1}} \dots g^{{\alpha_m} {\beta_m}}
F_{\alpha_1 \dots \alpha_m}F_{\beta_1 \dots \beta_m}.
\end{equation}

Here $R$ is the Ricci scalar of the metric $g_{\alpha \beta}$, $e^a_\alpha$
is the vielbein which squares to $g_{\alpha \beta}$ (and $e$ is its
determinant), $A_3$
is the three-form field discussed in the previous section,
and $\psi$ is the spin 3/2 gravitino.  We use the notation
\begin{eqnarray}
\hat \omega_{\alpha ab} &=& \omega_{\alpha ab} + \frac{1}{8}
\overline{\psi}^\beta \Gamma_{\beta \alpha ab \gamma} \psi^\gamma, \cr
\hat{F}_{\alpha \beta \gamma \delta} &=& F_{\alpha \beta \gamma \delta}
- 3 \overline{\psi}_{[\alpha} \Gamma_{\beta \gamma} \psi_{\alpha]}.
\end{eqnarray}
In the above, Greek letters ($\alpha, \beta,...$) denote spacetime indices
and Latin letters ($a,b,...$) denote internal indices.   The square
brackets $[...]$ indicate a completely antisymmetric sum over permutations
of the indices, divided by the number of terms.
Our conventions for spinors and
$\Gamma$-matrices are those of \cite{GSW}.  We will not state
them explicitly here as spinors only make appearances in this section and
section \ref{bps} and, in both cases, the details can be safely glossed
over.

This looks a little complicated, but let's
take a minute to sort through the various terms.
We'll begin with the least familiar part: the gravitino.
Since our attention here will be focused on classical solutions, we
will be able to largely ignore the gravitino.  The point
here is that the gravitino is a fermion and, due to
the Pauli exclusion principle, fermion fields do not have
semi-classical states of the same sort that bosonic fields do.
It is helpful here to think about the electron field as an
example.  There are, of course, states with a large number
of electrons that are well described by a classical charged
fluid.  However, because of the exclusion principle,
there are no semi-classical coherent states of the electron field
itself; i.e., no
states for which the dynamics is well described by a
{\it classical spinor field}.  In the same way, we might
expect that there are states of the gravitino field that
are well described by some sort of classical fluid, but we should
only expect the classical action (\ref{eleven}) to be
a good description of the dynamics when the gravitino field
vanishes.  Thus, we will set $\psi =0$ throughout most of
our discussion.  This is self-consistent at the classical level as setting $\psi =0$
in the initial data is enough to guarantee $\psi = 0$ for all time.

A study of (\ref{eleven}) shows that the dynamics of the solutions
for which $\psi =0$ can be obtained by simply setting $\psi$
to zero in the action.  This simplifies the situation
sufficiently that it is worth rewriting the action as:

\begin{equation}
\label{11B}
S_{\rm bosonic} = \frac{1} {2 \kappa_{11}^2 } \int d^{11} x (-g)^{1/2}
\bigl( R - \frac{1}{2} |F_4|^2 \bigr) - \frac{1}{12\kappa_{11}^2}
\int A_3 \wedge F_4 \wedge F_4.
\end{equation}
This sort of presentation, giving only the bosonic terms,
is quite common in the literature and
is sufficient for most solutions of interest\footnote{Typically,
an interesting bosonic solution in fact corresponds to
a quantum state in some supersymmetry multiplet.
The supersymmetry algebra can often be used to construct
spacetime solutions corresponding to other states in the multiplet in which
the fermions are excited.  See for example \cite{mult}.}.
Now that the gravitino has been set to zero, we see that
our action contains only three terms: the Einstein (scalar
curvature) term $R$, the Maxwell-like term $|F_4|^2$, and
the remaining so-called `Chern-Simons term.' \index{Chern-Simons term}

Note that, although the Chern-Simons term contains the gauge potential $A_3$, it is invariant under gauge transformations $A_3 \rightarrow A_3 + d\Lambda_2$, at least under `small' gauge transformations for which $\Lambda$ is a well-defined smooth 2-form and $\Lambda_2 \rightarrow 0$ sufficiently rapidly at any boundaries.  This term is very interesting, and turns out to significantly modify the picture outlined in section \ref{forms}.

Though we will be able to ignore these modifications for our purposes below, it is worth saying just a few words about them here.  To this end,  suppose that we couple a 3-form source $J_3$  to (\ref{11B}) via a term of the form $\int A_3 \wedge \star J_3$.  The $F_4$ equation of motion
is then of the form
\begin{equation}
\label{Jsource}
d(\star F_4 + (const) A_3 \wedge F_4) = \star J_3.
\end{equation}
suggesting that the relevant conserved 2-brane charge $Q_2$ is given by integrating $\star F_4 + (const) A_3 \wedge F_4$ over some closed 7-surface. The Bianchi identity on $F_4$ means that $Q_2$ is invariant under `small' gauge transformations though, like the action, $Q_2$ does change under the action of `large' gauge transformations where $\Lambda_2$ is not single-valued.  In fact, in units of the fundamental charge (discussed below), one can show that $Q_2$ shifts by an integer multiple of an associated 5-brane charge $Q_5$ defined by integrating $F_4$ over a 4-surface.  This feature allows M2-branes to end on M5-branes. See e.g. \cite{Town} for a review of this phenomenon and \cite{3Q} for further comments on charge in the presence of Chern-Simons terms.

The reason we can ignore the Chern-Simons term below is that we will consider
only relatively simple configurations of branes.  Specifically
we note that since $dF =0$, the
Chern-Simons contribution to (\ref{Jsource}) is
proportional to $F_4 \wedge F_4$.  Furthermore, the Chern-Simons term makes no contribution at all to the equation of motion for the metric.  Thus, whenever there are 4 or more
linearly independent vectors $k$ at each point such that
$k^\alpha F_{\alpha \beta \gamma \delta} =0$, we have $F_4 \wedge F_4 =0$
and the Chern-Simons term does not affect the equations of motion.
For the cases we consider below, this property is
satisfied as all of the non-vanishing components of $F$ will
lie in a subspace of dimension seven or less.

\subsection{Supersymmetry and BPS states}
\label{bps}

Some comments are now in order on the subject of supersymmetry, so
that we may introduce (and then use!) the concept of Bogomuln'yi-Prasad-Sommerfeld (BPS) states.
Again, I would like to begin with a few heuristics to provide a rough
perspective for students of 3+1 general relativity. We will see below
 that BPS solutions are closely related to extremal solutions;
in particular, to extremally charged solutions.
As a result, most of our intuition from extreme Reissner-Nordstr\"om
solutions carries over to
the general BPS case.

The setting for any discussion of BPS solutions is
the class of supergravity solutions which satisfy either asymptotically flat or asymptotically Kaluza-Klein boundary conditions.
We thus require the topology of the asymptotic region to be of the form $({\mathbb R}^n - \Sigma) \times Y$ for some compact set $\Sigma \subset {\mathbb R}^n$ and some homogeneous manifold $Y$,  though we will not specify further details of the boundary
conditions here.

In the setting of pure gravity, one would expect  such spacetimes
to exhibit asymptotic symmetries that correspond to the
Poincar\'e group in the appropriate number of
spacetime dimensions together with the symmetries of $Y$.  Some particular
solutions in this class will even have Killing vectors which make
some subgroup of the Poincar\'e group into an {\it exact} symmetry of the
spacetime; e.g. the rotation subgroup in spherically symmetric cases.

Now, supersymmetry \index{supersymmetry} is best thought of as an (anti-commuting)
extension of the diffeomorphism group.  Indeed, diffeomorphisms
form a subgroup of the supersymmetry gauge transformations and,
in the asymptotically flat setting just described, the
asymptotic Poincar\'e transformations will be a subgroup of
the asymptotic supersymmetry transformations.  Solutions that
are invariant under a subgroup of the supersymmetry transformations
containing non-trivial anti-commuting elements are said to have
a `Killing spinor' $\eta$ and are known as BPS (Bogomuln'yi-Prasad-Sommerfeld)
solutions. \index{Killing spinor}

It turns out that $\overline{\eta} \Gamma^\mu \eta$ is then a Killing vector field of the solution.  Furthermore, it is everywhere non-spacelike (i.e., timelike or null).\footnote{When the spacetime is asymptotically flat in all directions (i.e., the fields decay in an asymptotic region diffeomorphic to
${\mathbb R}^n$ minus some compact set), this Killing field is
in fact timelike, except perhaps on a Killing horizon \cite{GH}. The structure of the argument is very much
like the Witten proof of the positive energy theorem \cite{Witten}.}
Since for non-extreme black holes any stationary Killing field becomes spacelike behind the horizon,  it follows that any BPS black hole must be extreme.

Having oriented ourselves with this intuitive introduction, it
is now time to examine the details of the eleven-dimensional
supersymmetry transformations and their algebra.  The infinitesimal
supersymmetry transformations are in one-to-one correspondence with
Grassmann valued (Majorana\footnote{i.e., satisfying the reality
condition $\eta^* = B\eta$
where $B = \Gamma^3 \Gamma^5 ... \Gamma^9$ and $*$ denotes complex
conjugation.}) spinor fields $\eta(x)$.  The transformation
associated with $\eta$ is given by

\begin{eqnarray}
\label{susy}
\delta e^a_\alpha &=& \frac{1}{2} \overline \eta \Gamma^a \psi_\alpha, \cr
\delta A_{\alpha \beta \gamma} &=&  - \frac{3}{2}
\overline \eta \Gamma_{[\alpha \beta }\psi_{\gamma]},\cr
\delta \psi_\alpha &=&  D_\alpha(\hat{\omega}) \eta
+ \frac{\sqrt{2}}{(4!)^2}\bigl( \Gamma_\alpha^{abcd} - 8 e_\alpha^a \Gamma^{bcd}
\bigr) \eta \hat{F}_{abcd} \equiv \hat D_\alpha \eta,
\end{eqnarray}
where the last line defines the supercovariant derivative $\hat D_\alpha$
acting on the spinor $\eta$.

The details of the  supersymmetry
transformations are not particularly important
for our purposes.
What {\it is} important is the general structure.  Note that
the variation of the vielbein $e$ involves the gravitino $\psi$, but
then the variation of the gravitino involves the connection $\hat \omega$
which contains derivatives of the vielbein.  Similarly taking two
variations of the gauge field $A_3$, we find terms involving derivatives
of the gauge field.  As a result, the proper second
variations give just diffeomorphisms of the spacetime.

Recalling that the variation of $A_3$ contains $\psi$, we
also note that the first variation of the gravitino field
involves a derivative of the spinor $\eta$.  Thus, the second variation
of $A_3$ is something that involves the derivative of $\eta$.
With the proper choice of spinors $\eta$, one can construct a second
supersymmetry variation that gives just the usual gauge transformation
$A_3 \rightarrow A_3 + d \Lambda_2$ on the gauge field, where this $\Lambda_2$ is built from the relevant $\eta$'s.    Thus,
both diffeomorphisms and gauge transformations are in fact contained
in the spacetime supersymmetry algebra.  The supersymmetry algebra can
be thought of as a sort of `square root' of the diffeomorphism and gauge
algebras.   The fact that diffeomorphisms and gauge transformations
are expressed as squares leads to extremely useful positivity
properties.

While we will not need the details of the local supersymmetry algebra below,
it is useful to display the
algebra of the asymptotic supercharges. \index{supercharges}
Just as for diffeomorphisms and
gauge transformations, the asymptotic supersymmetries
lead, in the asymptotically flat context,
to conserved `supercharges.'  In fact, for the eleven dimensional
case, there are several relevant notions of the asymptotic algebra.
This is because there are interesting $p$-branes with several
values of $p$.  There are thus several interesting classes of
asymptotically Kaluza-Klein structures associated with different choices
of the homogeneous manifold $Y = {\mathbb R}^{11-p}$.

However, all of these algebras are rather similar.
If $Q$ is the generator of supersymmetry transformations,
so that the asymptotic versions of the
transformations above are generated by taking (super) Poisson
brackets with $Q \overline \eta$, then the algebra associated with
the $p$-brane case has the general form (see e.g. \cite{AGT})

\begin{equation}
\{Q_A^I,\overline{Q}^{JB}\}_+ = -2 P_\mu \Gamma^{\mu B}_A \delta^{IJ}
 - 2 i Z^{IJ} \delta_A^B,
\end{equation}
where we have used $A,B$ for the internal spinor indices.
Here, $P_\mu$ are the momenta per unit $p$-volume and $Z^{IJ}$
is an antisymmetric real matrix associated with the asymptotic
gauge transformations.  In
particular, the eigenvalues of $Z$ are of the form $\pm i q$
where $q$ is the appropriately normalized
charge carried by the $p$-brane.  Our notation
reflects the fact that it is natural to split the SUSY generator Q,
which is an eleven dimensional Majorana fermion, into a set
of $(11-p)$ dimensional fermions $Q^I$.   Thus, the indices $A,B$
take values appropriate to spinors in $(11-d)$ dimensions.

The most important property of this algebra is that it implies
the so-called BPS bound on masses and charges.  To get an idea of
how this arises,
recall that while $\overline {Q} Q$ is a Lorentz invariant, it is
$Q^\dagger Q$  that is a positive definite operator.  Thus, a positivity
condition should follow by writing the algebra in terms of $Q^\dagger$ and
$Q$.  For simplicity, let us also choose an asymptotic
Lorentz frame such that energy-momentum of the spacetime is aligned with
the time direction: $P_\mu = T \delta_{\mu0}$, where $T$ is the
brane tension, or mass per unit $p$-volume\footnote{Simple relativistic branes have a fixed value of $T$. As a result, stretching the brane over an additional $p$-volume $V_p$ increases the energy by $TV_p$ and requires a  corresponding amount of work.  Thus $T$ is also the tension in the more familiar sense of the force-density (per unit $(p-1)$-volume) required to stretch the brane.}.  The algebra then
takes the form

\begin{eqnarray}
\label{simp}
\{Q_A^I,Q^{\dagger JB}\}_+ &=& 2T \delta^{IJ} \delta_A^B + 2i Z^{IJ}
\Gamma^{0B}_A.
\end{eqnarray}

It is useful to adopt the notation of quantum mechanics, even though
we are considering classical spacetimes.  Thus, we describe a spacetime
by a state $|\psi\rangle$ and we let the generators $Q$ act on that
state as $Q|\psi\rangle$.   Contracting the above relation (\ref{simp})
with $\eta_{JB}$ and $\eta^{\dagger A}_I$ for a set of spinor
fields $\eta_I$, taking
the expectation value in any state, and using the positivity of the
inner product and the fact that the eigenvalues of $\Gamma^{0B}_A$ are
$\pm1$ yields the relation
\begin{equation}
\label{BPS}
T \ge  |q|.
\end{equation}
See \cite{GHT} for a full classical supergravity
derivation in the context of magnetic charge in eleven dimensions and
\cite{GH} for a complete derivation in classical
$N=2$ supergravity in four dimensions.  See also \cite{alg}
for details of the argument above  in the four dimensional context.

This is the BPS bound.  \index{BPS bound} A spacetime in which this
bound is saturated is called a BPS spacetime and the corresponding
quantum states are known as BPS states.  \index{BPS states} Note that, from our above
argument, a state is BPS only if it is annihilated by one of the
supersymmetry generators; that is, if the spacetime is invariant under
the transformation (\ref{susy}) for some spinor $\eta$.
The converse is also true; any asymptotically flat
spacetime which is invariant under
some nontrivial supersymmetry transformation is BPS.
Given a solution $s$ and a spinor $\eta$
for which the transformation (\ref{susy}) vanishes on $s$, one says
that $\eta$ is a {\it Killing spinor} of $s$.\index{Killing spinor}
Since the gravitino $\psi$ vanishes for a bosonic solution, in this
context we see from (\ref{susy}) that $\eta$ is
a Killing spinor whenever it is supercovariantly constant; i.e when it
satisfies $\hat D_\alpha \eta = 0$.

The bound (\ref{BPS}) is reminiscent of the extremality bound for
Reissner-Nordstr\"om black holes. \index{extremal limit}  It turns out that the
relationship is a strong one.
Given the similarity of eleven dimensional supergravity to
Einstein-Maxwell theory, it will come as no surprise that
there is a supergravity theory in 3+1 dimensions that
contains Einstein-Maxwell theory, together with a few extra fields.
When the extra fields vanish on an initial slice, they remain zero
for all time.  Thus,
Einstein-Maxwell theory is a `consistent truncation' of
the supergravity.  In this context, the BPS bound and the
extremality bound for charge coincide when there is no angular
momentum.  Thus, any asymptotically
flat solution of Einstein-Maxwell theory with extremal charge and vanishing
angular momentum can be lifted to a BPS solution of the supergravity.

In general, any BPS black hole solution will be extremal, though
the converse is not always true.  An important example occurs
in four dimensions where all BPS states must have zero angular momentum.
Thus, the 3+1 extreme Kerr solution is not BPS.

Now, in familiar 3+1 Einstein-Maxwell gravity, we are used to thinking
of extreme black holes as being some sort of marginal and perhaps
unphysical case.  Indeed, it is an important part of black hole
thermodynamics that one cannot by any finite (classical)
process transform a non-extreme black hole into an extreme black hole.
Moreover, a real astrophysical black hole will quickly lose its
charge due to interactions with the interstellar medium.  Even in a pure
vacuum, quantum field theory effects in the real world cause black holes
to loose their charge and to evolve toward neutral black holes.
However, this last statement is a consequence of the large
charge to mass ratio of the electron.  Due to the BPS bound discussed above, objects like the familiar
electron do not exist in a theory with sufficient supersymmetry.  As a result, BPS black holes
do not discharge.  Instead,
non-extreme black holes decay {\it toward}
extremality through the emission of Hawking radiation and (assuming that black holes have only a finite number of internal states)
any non-extreme black hole will decay
to an extreme black hole in a very large but finite time.
Thus, extreme black holes are of central importance in understanding
supersymmetric theories as they represent stable `ground states'
for black holes.

Now that we have come to terms with supersymmetry, we will
proceed to ignore fermions completely in the sections below.

{\bf Executive Summary:} If you decided to skip over the material involving spinors, the key point is that theories with enough supersymmetry have a so-called BPS bound.  This means that
(in appropriate units) the mass per unit volume of a $p$-brane must be greater or equal to than the associated $A_{p+1}$-form charge.  Solutions that saturate this bound are called BPS solutions and have special properties.  In particular, any BPS black hole or brane is extremal.

\section{M-branes: The BPS solutions}
\label{Mbr}

Although we wish to focus on the eleven dimensional case, supersymmetry and supergravity can also be considered in less
than eleven dimensions.  For example, in 3+1 dimensions any asymptotically flat solution of Einstein-Maxwell
theory with extremal charge and zero angular momentum can be lifted to a BPS solution
of 3+1 supergravity.  But this is just the class
of Majumdar-Papapetrou solutions \cite{MP}, which
consist of some number of extreme Reissner-Nordstr\"om black
holes in static equilibrium.  \index{Majumdar-Papapetrou solutions} Since the Majumdar-Papapetrou solutions
are a more familiar analogue of the eleven dimensional M-brane
solutions which we wish to discuss, we
present a brief review in
section \ref{MPsols} as an introduction to the world of M-branes.  {\sf} This material overlaps with and extends the discussion of Chapter 1.  We then examine the
M-branes themselves in section \ref{branes}.

\subsection{The 3+1 Majumdar-Papapetrou solutions}
\label{MPsols}

Recall that the Reissner-Nordstr\"om solution with
mass $M$ and charge $Q$
takes the form

\begin{equation}
\label{RNmetric}
ds^2 = - (1 - \frac{2GM}{R} + \frac{GQ^2}{R^2}) dt^2 +
\frac{1}{1 - \frac{2GM}{R} + \frac{GQ^2}{R^2}} dR^2 + R^2 d\Omega^2_2,
\end{equation}
with $R$ the usual Schwarzschild radial coordinate, $t$ the Killing
time, and $d\Omega^2_2$ the metric on the unit two-sphere.
Here, $Q$ and $M$ are the charge and mass of the black hole, with
$Q$ measured in units of $\sqrt{ (mass)(length)}$ as is natural in classical
mechanics with $c=1$ but $G\neq 1$; i.e., so that Coulomb's law is $F= Q^2/r^2$.  The factors of Newton's
constant $G$ have been left explicit for consistency
with the rest of this exposition.  The extremal situation
is $GM^2 = Q^2$ and, in this case, the solution is controlled
by a single length scale $r_0 =  GM = {\sqrt G}Q$.
The metric simplifies to take the form

\begin{equation}
\label{XRNmetric}
ds^2 = - (1 -  {{r_0}/R})^2 dt^2 +
(1 - {{r_0}/R})^{-2} dR^2 + R^2 d\Omega^2_2.
\end{equation}

Let us take this opportunity to recall that the horizon at $r=r_0$ lies an infinite proper distance away from any $r > r_0$ along any surface of constant $t$.  Since the size of the sphere is approximately constant ($r_0$) over the entire region near the horizon one says that an extreme black hole has ``an infinite throat."   In fact, the region near the horizon is just the so-called Bertotti-Robinson universe, 2-dimensional anti-de Sitter space (AdS${}_2$) times  $S^2$.  A simple way to see this is to let $z = r_0(1-r_0/R)^{-1}$ and to expand the metric in powers of $1/z$.  One finds
\begin{equation}
\label{BT}
ds^2 = \frac{- dt^2 + dz^2}{z^2} + r_0^2 d\Omega^2_2 + O(z^{-4}).
\end{equation}
The leading terms shown give precisely AdS${}_2 \times S^2$ in so-called Poincar\'e coordinates.

We now return to the full metric (\ref{XRNmetric}) and change to so-called isotropic coordinates in which the spatial part of the metric is conformally flat. Let
$r = R- r_0$, so that the horizon lies at $r=0$.  Introducing
the Cartesian coordinates $x^i$ as usual on ${\bf R}^3$, we have

\begin{equation}
\label{isoRN}
ds^2 = - f^{-2} dt^2 +
f^2 \sum_{i=1}^3 dx^i dx^i,
\end{equation}
where $f = 1 + r_0/r$.  Similarly, the electro-magnetic potential
is given by $A_t =  f^{-1}$ with the spatial components
of $A$ vanishing.  As the function $f$ satisfies
Poisson's equation with a delta function source,

\begin{equation}
\label{1stP}
\partial^2_x f := \sum_{i=1}^3 \partial_i
\partial_i f = - 4\pi  \delta^{(3)}(x),
\end{equation}
the solution for the extreme black hole takes a form
similar to that seen in electrodynamics (except that the
Poisson equation is for the {\it inverse} of the electrostatic
potential).  Note that the relevant differential
operator is the Laplacian on a {\it flat} three-space and not
the one directly defined by the metric.  Such differential
operators will often appear below, and we will use the
convention that $\partial_x^2$ will always denote the flat-space
Laplacian associated with the coordinates $x$. Similarly,
we will write $dx^2 := \sum_{i} dx^i dx^i$.

The analogy with electrostatics is quite strong.
The above metric (\ref{isoRN}) and the associated electric field
 define the class of Majumdar-Papapetrou solutions \cite{MP}.
These are, in general, solutions of the Einstein-Maxwell
system coupled to so-called extremal dust.  This dust is characterized by the property that, when two grains of dust are at rest,
their electrostatic repulsion is exactly sufficient to balance their
gravitational attraction and they remain at rest.
Modulo the conditions below,
any choice of the function $f$ in (\ref{isoRN}) yields a static solution
of the field equations corresponding to some distribution of
this dust.  For an asymptotically flat solution, we should
take $f$ to be of the form $1 + Q/r$ near infinity.  The one
restriction on $f$ is that
$\rho = -\frac{1}{4\pi} \nabla^2 f$ must be everywhere positive.
In particular, we
will take it to be of the form
$\rho_0 + \sum_{k =1}^N r_k \delta(x-x_k)$ where $\rho_0$
is continuous.  The density (defined with respect to the
Cartesian coordinate system $x_i$) of extremal dust is given
by $\rho_0$ and each delta function will result in the presence
of an extremal black hole.

Extremality is quite important for the simple form of this class of solutions.
It is only in the extreme limit that the repulsion induced by the
electric charge can `cancel' the gravitational attraction so that the
solution can remain static. If one adds any additional energy to the
solution, the non-linearities of gravity become more directly manifest.

Note that the source in (\ref{1stP}) lies
at the origin of the $x$-coordinates; i.e., at the horizon
of the black hole.  However, since the horizon of the black hole
is in fact not just a single point in space, $x=0$ is clearly
a coordinate singularity.  This means that although the support of the
delta function lies at $x=0$, this should not be interpreted
as the location of the black hole charge.  Rather, the role of
this delta function is to enforce a boundary condition on the
electric flux emerging from the black hole so that the hole
does indeed carry the proper charge.

Of course, in 3+1 dimensions, we can also have magnetically
charged black holes.  In fact, we can have dyons, carrying
both electric and magnetic charge.  The corresponding
extremal solutions are given directly by electro-magnetic
duality rotations
of the above solution.

For future reference we note that there is a similar
set of solutions in 4+1 dimensions, though black holes
in five dimensions can carry only electric charge.
They take the form
\begin{equation}
\label{5RN}
ds^2 = - f^{-2} dt^2 +
f \sum_{i=1}^4 dx^i dx^i = -f^{-2} dt^2 + f dx^2,
\end{equation}
where $\partial_x^2 f = -2 \Omega_3 (\rho_0 + \sum_{k=1}^N r_k^2
\delta(x-x_k))$, $\Omega_3$ is the volume of the unit
three-sphere, $\rho_0$ is the
charge per unit $d^4x$ cell, and $r_k$ is a length scale parametrizing the charge of the $k$-th
extremal black hole.  The fact that $r_k^2$, as opposed to $r_k$,
appears as the source
reflects the fact that the
fundamental solution of Poisson's equation in four dimensions is of the form
$r^{-2}$.

As we have
already commented, there is a coordinate singularity at
the black hole horizon.  Thus, the isotropic form of the metric
does not allow us to see to what extent the black hole, or even
the horizon, is non-singular.  However, if the black hole
is to have a smooth horizon, then a necessary condition
is that the horizon have non-zero (and finite) area.
That this is true of the above metrics is easy to read
off from (\ref{isoRN}) and (\ref{5RN}) by realizing that the
divergence of $f^2$ or $f$ cancels the $r^2$ factor
that arises in writing $dx^2 = dr^2 + r^2 d\Omega^2$ in
spherical coordinates.  While this is certainly not
a sufficient condition for smoothness of the horizon, it
will serve as a useful guide below.\footnote{Interestingly, while the 3+1 Majumdar-Papapetrou  solutions (\ref{isoRN}) are smooth (and in fact analytic \cite{HHMP}), the 4+1 solutions (\ref{5RN}) are not \cite{GHT,Welch}. In even higher dimensions the corresponding solutions have curvature singularities on the would-be horizon \cite{GCHR}.}

\vbox{
\centerline{\begin{picture}(1331,2566)(3208,-1944)
%  METADATA <id>1</id>
{\color[rgb]{0,0,0}\thinlines
\put(3295,-661){\circle*{150}}
}%
%  METADATA <id>2</id>
{\color[rgb]{0,0,0}\put(3295,-1861){\circle*{150}}
}%
%  METADATA <id>3</id>
{\color[rgb]{0,0,0}\put(3295,539){\circle*{150}}
}%
%  METADATA <id>4</id>
{\color[rgb]{1,0,0}\multiput(3295,-661)(3.75000,-7.50000){21}{\makebox(1.6667,11.6667){\tiny.}}
\multiput(3370,-811)(-4.68750,-7.03125){33}{\makebox(1.6667,11.6667){\tiny.}}
\multiput(3220,-1036)(4.68750,-7.03125){33}{\makebox(1.6667,11.6667){\tiny.}}
\multiput(3370,-1261)(-4.68750,-7.03125){33}{\makebox(1.6667,11.6667){\tiny.}}
\multiput(3220,-1486)(4.68750,-7.03125){33}{\makebox(1.6667,11.6667){\tiny.}}
\multiput(3370,-1711)(-3.75000,-7.50000){21}{\makebox(1.6667,11.6667){\tiny.}}
}%
%  METADATA <id>5</id>
{\color[rgb]{1,0,0}\multiput(3295,539)(3.75000,-7.50000){21}{\makebox(1.6667,11.6667){\tiny.}}
\multiput(3370,389)(-4.68750,-7.03125){33}{\makebox(1.6667,11.6667){\tiny.}}
\multiput(3220,164)(4.68750,-7.03125){33}{\makebox(1.6667,11.6667){\tiny.}}
\multiput(3370,-61)(-4.68750,-7.03125){33}{\makebox(1.6667,11.6667){\tiny.}}
\multiput(3220,-286)(4.68750,-7.03125){33}{\makebox(1.6667,11.6667){\tiny.}}
\multiput(3370,-511)(-3.75000,-7.50000){21}{\makebox(1.6667,11.6667){\tiny.}}
}%
%  METADATA <id>6</id>
{\color[rgb]{0,0,1}\put(3287,539){\line( 1,-1){600}}
\put(3887,-61){\line(-1,-1){600}}
\put(3287,-661){\line( 1,-1){600}}
\put(3887,-1261){\line(-1,-1){600}}
}%
%  METADATA <id>7</id>
{\color[rgb]{0,0,0}\put(4485,528){\line(-1,-1){600}}
\put(3885,-72){\line( 1,-1){600}}
\put(4485,-672){\line(-1,-1){600}}
\put(3885,-1272){\line( 1,-1){600}}
}%
%  METADATA <id>8</id>
\put(4347,-1192){\makebox(0,0)[lb]{\smash{{\SetFigFont{17}{20.4}{\rmdefault}{\mddefault}{\updefault}${\cal I}^-$}}}}
%  METADATA <id>9</id>
\put(4374,-460){\makebox(0,0)[lb]{\smash{{\SetFigFont{17}{20.4}{\rmdefault}{\mddefault}{\updefault}${\cal I}^+$}}}}
\end{picture}%
}
\centerline{Figure 11.2 Conformal diagram for extreme Einstein-Maxwell
black holes.}}

\medskip

For completeness, we display the conformal
diagrams for the above solutions in Fig. 11.2.  The 3+1 and 4+1 cases are identical
except for the dimension of the (suppressed) spheres of symmetry.
Here ${\cal I^+, I^-}$ denote the past and future null infinities of a particular
asymptotic region and the wavy line on the right
denotes the (timelike) singularity.
The black circles mark the `internal infinities.'  These points
lie at an infinite affine parameter along any geodesic (spacelike,
timelike, or null) from the interior.

However, Fig. 11.2 does not tell the entire story.  For charged non-extreme black holes it is known that, while the exterior solution is quite stable, the simple analytic textbook Reissner-Nordstr\"om solution is in fact unstable near the inner horizon.  Perturbations transform the would-be inner horizon into a curvature singularity, see e.g. \cite{PI,Dafermos,BradySmith}.  Now, recall that the inner horizon coincides with the event horizon in the extreme limit.  Though there are some subtleties, this turns out to imply that extreme black holes (produced, say, by the emission of Hawking radiation from non-extreme black holes) have a singularity lurking just below their event horizon \cite{extremes}.  Specifically, in the limit where an observer enters the black hole long after the black hole was formed, no proper time passes for that observer between the horizon and the singularity. We say that such black holes have an `effective spacetime diagram' with a singular horizon given by Fig. 11.3.
We expect that the same is true of the higher dimensional extreme black holes discussed below, where by black {\it hole} we mean that the horizon is compactly generated.  Black branes with noncompactly-generated horizons can be quite different due to the extra need to impose boundary conditions in the directions along the branes.

\vbox{
\begin{center}
\begin{picture}(1456,1374)(3355,-1273)
%  METADATA <id>1</id>
{\color[rgb]{0,0,0}\thinlines
\put(3438,-661){\circle*{150}}
}%
%  METADATA <id>4</id>
{\color[rgb]{1,0,0}\multiput(3811,-661)(2.00000,-12.00000){2}{\makebox(1.6667,11.6667){\tiny.}}
}%
%  METADATA <id>5</id>
{\color[rgb]{1,0,0}\multiput(4018,-12)(-1.65625,-8.28125){33}{\makebox(1.6667,11.6667){\tiny.}}
\put(3965,-277){\line(-5,-1){265}}
\multiput(3700,-330)(-1.66226,-8.31130){33}{\makebox(1.6667,11.6667){\tiny.}}
\multiput(3647,-596)(-7.89512,-3.15805){25}{\makebox(1.6667,11.6667){\tiny.}}
}%
%  METADATA <id>6</id>
{\color[rgb]{0,0,1}\put(3437,-661){\line( 1,-1){600}}
}%
%  METADATA <id>7</id>
{\color[rgb]{0,0,0}\put(4045,-61){\line( 1,-1){600}}
\put(4645,-661){\line(-1,-1){600}}
}%
%  METADATA <id>8</id>
\put(4654,-1232){\makebox(0,0)[lb]{\smash{{\SetFigFont{17}{20.4}{\rmdefault}{\mddefault}{\updefault}${\cal I}^-$}}}}
%  METADATA <id>9</id>
\put(4466,-318){\makebox(0,0)[lb]{\smash{{\SetFigFont{17}{20.4}{\rmdefault}{\mddefault}{\updefault}${\cal I}^+$}}}}
\end{picture}%

Figure 11.3 Effective conformal diagram for perturbed extreme Einstein-Maxwell
black holes as seen by late-time observers.
\end{center}
}

\subsection{Brane solutions in eleven dimensions}
\label{branes}

There are four (basic) solutions of eleven dimensional
supergravity that are of particular importance in string/M-theory.
They are known as the (eleven-dimensional)
Aichelburg-Sexl metric \cite{AS}, the
M2-brane \cite{M2sol}
(electrically charged under $A_3$), the M5-brane (magnetically
charged under $A_3$), and the eleven dimensional
version of the Kaluza-Klein monopole \cite{Rafael,GP}.
Although it may not be obvious from the names, all four of the basic
solutions are associated with branes in string/M-theory.
Below, we discuss only the extreme versions of these basic solutions, which turn out to be BPS.
In particular they
each have 16 Killing spinors, preserving half of the supersymmetry.
The non-extremal forms of the M-branes may be found in, e.g., \cite{GHT}.

It is sometimes said that an arbitrary BPS solution can
be built from these basic solutions.  The point here is that
the above `basic' solutions are in one-to-one correspondence
with the types of charge present in eleven dimensional
supergravity.  Since the charges are additive, one is tempted
to say that any solution with arbitrary amounts of the various charges
can be built up by `combining' these basic solutions.  We will see
below that certain simple solutions carrying multiple charge
are in fact built from the basic solutions in
a simple way.  However, there is as yet no known method
for writing down a general BPS solution at all, much less
in terms of the basic solutions.

Let us begin with the BPS M2- and M5-branes as these are straightforward supergravity
analogues of the extreme Reissner-Nordstr\"om black holes discussed above\footnote{Interestingly, the
global structure of the {\it non-extreme}
M2- and M5-brane solutions is much like that
of the Schwarzschild black hole, as opposed to that
of non-extreme Reissner-Nordstr\"om.  In particular, there is no
inner horizon and the singularity is spacelike as opposed to timelike.}.
There is a corresponding notion of
isotropic coordinates in which the multi-black hole
solutions are given by solving a flat space Poisson equation
with delta-function sources.  The solutions of this Poisson equation
are typically
denoted $H_2$ for the M2-brane and $H_5$ for the M5-brane
and are referred to as `harmonic' functions.  The details are different
for the two branes, but both should seem quite familiar from our review
of the Majumdar-Papapetrou solutions.

For the M2-brane, \index{M2-brane} we introduce a set of three coordinates
$x_{\parallel}$ which should be thought of as labeling
the directions along the brane, and a set of eight coordinates
$x_\perp$ which should be thought of as labeling the
directions orthogonal to the brane.  As one of the
$x_{\parallel}$ directions is the time direction,
we define $dx^2_{\parallel} = - (dx_\parallel^0)^2 +
(dx_\parallel^1)^2 + (dx_\parallel^2)^2$.
The solution takes the form:

\begin{eqnarray}
\label{M2}
A_3 &=&  H_2^{-1} dt \wedge dx_{\parallel,1} \wedge dx_{\parallel,2}
\cr
ds^2 &=&  H_2^{-2/3} dx_{\parallel}^2 + H_2^{1/3} dx_{\perp}^2,
\end{eqnarray}
with $\partial_\perp^2 H_2$ equal to a sum of delta-functions.
Note that, near the delta function source, $H_2$ will diverge
like $r^{-6}$, where $r$ is the $x_\perp$ coordinate distance from the
source.  As a result, $H_2^{1/3}$ diverges like
$r^{-2}$, and the sphere at the horizon will
have non-zero (finite) area.   This suggests that the
horizon of at least a single BPS M2-brane is smooth, and a careful
investigation \cite{GHT} does indeed show that this is
the case.  In fact, by a direct analogue of our discussion for 3+1 extreme Einstein-Maxwell black holes, one can show that the region near the horizon is just AdS${}_4 \times S^7.$

This is rather interesting, as the extremal
limits of black branes in lower dimensional supergravity theories tend, because
of the dilaton, to have singular horizons.  The
global structure of the M2-brane is in fact much like
that of the extreme Reissner-Nordstr\"om black holes discussed
above.  The conformal diagram is just that of Fig. 11.1,
except that each point on the diagram now represents
a surface with both the topology and metric of
${\bf R}^2 \times S^7$ instead of just a sphere.

For the M5-brane, \index{M5-brane} we introduce a set of six coordinates
$x_{\parallel}$ along the brane, and a set of five coordinates
$x_\perp$ orthogonal to the brane.  Again, the $x_\parallel$
directions include the time $t$.
The solution takes the form:

\begin{eqnarray}
\label{M5}
dA = F &=& - \frac{1}{4!} \partial_{x_\perp^i} H_5 \epsilon^{i}{}_{jklm}
dx^j \wedge dx^k \wedge dx^l \wedge dx^m
\cr
ds^2 &=&  H_5^{-1/3} dx_{\parallel}^2 + H_5^{2/3} dx_{\perp}^2,
\end{eqnarray}
with $\partial_\perp^2 H_5$ equal to a sum of delta-functions.
The different form of the gauge field as compared with
(\ref{M2}) is associated with the
fact that this solution carries a magnetic charge instead of
an electric charge.  Now the field $H_5$ diverges
at a delta-function source as $r^{-3}$, so that $H_5^{2/3}$
diverges like $r^{-2}$ and again the area of the spheres
is finite at the horizon.  Once again, a detailed
study shows that the horizon is completely smooth and, as one might guess, the near-horizon geometry is AdS${}_7 \times S^4$.

\vbox{\centerline{
\begin{picture}(2022,2206)(2826,-4655)
%  METADATA <id>1</id>
\thinlines
{\color[rgb]{0,0,1}\put(3548,-2543){\line( 1,-1){600}}
\put(4148,-3143){\line(-1,-1){600}}
\put(3548,-3743){\line( 1,-1){600}}
\put(4148,-4343){\line(-1,-1){300}}
}%
%  METADATA <id>2</id>
{\color[rgb]{0,0,0}\put(4745,-2535){\line(-1,-1){600}}
\put(4145,-3135){\line( 1,-1){600}}
\put(4745,-3735){\line(-1,-1){600}}
\put(4145,-4335){\line( 1,-1){300}}
}%
%  METADATA <id>3</id>
{\color[rgb]{0,0,0}\put(3564,-2535){\line(-1,-1){600}}
\put(2964,-3135){\line( 1,-1){600}}
\put(3564,-3735){\line(-1,-1){600}}
\put(2964,-4335){\line( 1,-1){300}}
}%
%  METADATA <id>4</id>
\put(4702,-4141){\makebox(0,0)[lb]{\smash{{\SetFigFont{17}{20.4}{\rmdefault}{\mddefault}{\updefault}${\cal I}^-$}}}}
%  METADATA <id>5</id>
\put(4583,-3382){\makebox(0,0)[lb]{\smash{{\SetFigFont{17}{20.4}{\rmdefault}{\mddefault}{\updefault}${\cal I}^+$}}}}
%  METADATA <id>8</id>
\put(2841,-3615){\makebox(0,0)[lb]{\smash{{\SetFigFont{17}{20.4}{\rmdefault}{\mddefault}{\updefault}${\cal I}^-$}}}}
%  METADATA <id>10</id>
\put(2966,-2816){\makebox(0,0)[lb]{\smash{{\SetFigFont{17}{20.4}{\rmdefault}{\mddefault}{\updefault}${\cal I}^+$}}}}
%  METADATA <id>11</id>
\put(4064,-3817){\makebox(0,0)[lb]{\smash{{\SetFigFont{17}{20.4}{\rmdefault}{\mddefault}{\updefault}A}}}}
%  METADATA <id>12</id>
\put(3427,-3216){\makebox(0,0)[lb]{\smash{{\SetFigFont{17}{20.4}{\rmdefault}{\mddefault}{\updefault}B}}}}
%  METADATA <id>13</id>
\put(3427,-4491){\makebox(0,0)[lb]{\smash{{\SetFigFont{17}{20.4}{\rmdefault}{\mddefault}{\updefault}B}}}}
%  METADATA <id>14</id>
\put(4064,-2767){\makebox(0,0)[lb]{\smash{{\SetFigFont{17}{20.4}{\rmdefault}{\mddefault}{\updefault}A}}}}
\end{picture}%
}
\centerline{Figure 11.4  Conformal diagram for the extreme M5-brane.}}

\medskip

The surprise \cite{GHT} is that this solution turns out to have no singularities at all, even inside the horizon!  Its
conformal diagram is thus rather different from those we have
encountered so far and is shown in Fig.  11.4.    The regions
marked A and B below (`in front of' and `behind') the horizon
are exactly the same.  In familiar cases, the singularity
theorems guarantee that something of this kind does not
occur: compact trapped surfaces imply a singularity
in their future \cite{HawkingEllis}.  However,
the fact that we deal with a black brane, and not
a black hole, means that the trapped surfaces are not in
fact compact.  The point here is that the horizon
is extended in the $x_\parallel$ directions.  What happens
when the solution is toroidally compactified by making
identifications in  the $x_\parallel$ coordinates is an
interesting story that will be discussed below.

The remaining two solutions have the interesting property
of being BPS despite the fact that the gauge field $A_3$
is identically zero.  This is not really a contradiction to
the condition of extremality when one notes (see section \ref{SUGRAVKK})
that under Kaluza-Klein reduction a momentum can act like
a charge.   Another useful perspective results from recalling that
two parallel beams of light (or two parallel gravitational waves) do not
interact gravitationally.  The same is true for any null particles.
Thus, one may say that spatial components
of the momentum (of like sign) provide a gravitational repulsion and that the case
of null momentum is like the case of extremal charge, where this
repulsion just exactly balances the gravitational attraction due to the
energy of the particles.

The Aichelberg-Sexl metric carries just such a
null momentum.
This solution was originally constructed \cite{AS} (in 3+1
dimensions) by boosting a Schwarzschild solution while
rescaling its mass parameter $M$ in order to keep the total
energy $E$ finite in some asymptotic frame.  This explains
the null momentum of the resulting solution.  It too can
be described in terms of a `harmonic function' $H_{AS}$.
The Aichelberg-Sexl metric  may be thought of as the gravitational
field of a null {\it particle}, such as a graviton or a quantum of the
$A_3$ field in the
short wavelength (WKB) approximation.  We introduce a time
coordinate $t$, a coordinate $z$ in the direction of motion
of the particle, and a set of nine additional coordinates $x_\perp$.
In isotropic coordinates, the solution takes the form
\begin{eqnarray}
\label{Mwave}
ds^2 = - dt^2 + dx_\perp^2 +dz^2 +  (H_{AS}-1) (dt-dz)^2
\end{eqnarray}
where $H_{AS}(x_\perp)$ is a solution of $\partial_\perp^2 H_{AS} =
-7 \Omega_9 \rho$,
where $\rho$ is again a source and $\Omega_9$ is the volume of the unit
9-sphere.  When $\rho$ is
a delta-function, this solution is in fact singular at
the source.

Let us now turn to the Kaluza-Klein monopole.
This solution was originally constructed \cite{Rafael,GP} by using
the fact that the metric product of any two Ricci
flat spaces is Ricci flat.  Thus, one can make a static
solution of 4+1 Einstein gravity out of any solution to
four-dimensional, Euclidean gravity.  Such a solution is
Ricci flat, so the metric product with a line is also
Ricci flat.  The metric product of Euclidean Taub-NUT
space \cite{NUT} with a line gives the 4+1 Kaluza-Klein monopole.
The eleven dimensional solution of interest here is simply
the metric product of Euclidean Taub-NUT space with
a 6+1 Minkowski space.  Recall that the Taub-NUT
solution is not asymptotically flat.  Rather, it is asymptotically
flat in three directions (to which we assign coordinates
$x_{\perp}$) and the fourth direction (which we will call $\theta$)
is an angular coordinate for which the associated $S^1$
twists around the two-sphere to make a non-trivial asymptotic
structure.
Introducing coordinates $x_\parallel$
on the 6+1 Minkowski space, the solution takes the form:
\begin{equation}
\label{11KK}
ds^2 = dx_\parallel^2 + H_{KK} dx_\perp^2 + H^{-1}_{KK} (d\theta
+ a_i dx^i_\perp)^2
\end{equation}
with, of course, vanishing 3-form potential $A_3$.  Again, $H_{KK}$ satisfies
an equation of the form $\partial_\perp^2 H_{KK}= - 4 \pi \rho$ and
$a_k$ is determined from $H_{KK}$ via
$\partial_{x_\perp^i} H_{KK} = \epsilon_{ijk} \partial_{x_\perp^j} a_k$.
As usual, we find a coordinate singularity at the location of
the delta-function sources.

But the story of this singularity
is just that of Taub-NUT space.  Suppose that $\theta$ is periodic with
period $L$.  Then the spacetime is in fact smooth
in the neighborhood of a `source' of the form
$\rho= \frac{L}{4 \pi} \delta^{(3)}(x_\perp)$; in this case,
(\ref{11KK}) actually represents a smooth geodesically complete
solution to the {\it source-free} 10+1 Einstein equations.  A
source of this sort
is referred to as a monopole of unit charge.  A multi-center
solution\footnote{One with several delta-function sources.}
is smooth whenever each separate center has this charge.
Now, if we take a limit of a multi-center solution in which
several of the centers coalesce into a single center with charge
greater than one, the resulting spacetime has a timelike
singularity at the source.  However, this singularity (with an integer
$n$ number of units of the above fundamental charge) has a particularly
simple form. It is a quotient of flat space, and in this
sense it is a higher dimensional  version of a conical
singularity.

A favorite topic to include in discussions of black holes
is that of black hole entropy.  It is therefore natural to ask about
the entropy of the branes that we have discussed above.  Similar thermodynamic
arguments hold for black branes as for black holes, suggesting that
one should associate an entropy of $A/4G_{11}$ with such objects where
$A$ is the area (volume) of the horizon and
$1/16\pi G_{11} = 1/2 \kappa_{11}^{2}$
is the
coupling constant that stands in front of the supergravity action.
However, the Kaluza-Klein monopole and Aichelberg-Sexl metric
have no Killing horizons and so presumably
carry no such entropy.

The M2- and M5-branes are a bit more subtle.
On the one hand, their horizons are homogeneous surfaces that are non-compact.
As such, one might be tempted to assign them infinite entropy.
Some further
insight into the issue is gained by using the fact that the solutions
are invariant under translations in the spatial $x_{\parallel}$ coordinates
to make toroidal identifications and compactify the horizons.
We can then calculate the horizon area and, because the norms
of the Killing fields $\partial_{x_\parallel}$ vanish on the horizon,
the result is zero.  Thus, at least when compactified in this way,
the M2- and M5-branes have vanishing horizon volumes and again carry no entropy.  This is another sense
in which such solutions are `basic.'

One might guess that there is something singular about the zero-area horizons
of the compactified M2- and M5-branes.  However, since those solutions
were constructed by making discrete identifications of spacetime with
smooth horizons, the curvature and field strength cannot diverge
at the zero-area horizon.  It turns out that
the situation is essentially the same as that which
arises \cite{BTZ} when $AdS_3$ is identified to make the $M=0$
BTZ black hole.  The
initially spacelike Killing
fields $\partial_{x_\parallel}$ become null on the horizon but also have
fixed points.  Thus, the horizon of the compactified
solution has both closed null curves and a `Lorentzian conical singularity.'

\subsection{Brane engineering}
\label{smear}

Before leaving eleven dimensions, a few words are in order
on two of the basic techniques in `Brane Engineering,' constructing
new brane solutions from old.  The particular techniques
to be discussed are known as smearing and combining
charges.  Together, they will allow us to build BPS black branes with finite entropy.

Smearing \index{smearing} is particularly straightforward.  It is based
on the observation that each type of `basic' solution
above is related to the solution of a linear differential
equation.  Using a delta-function source gives a solution
which preserves some set of translation symmetries (in the
$x_{\parallel}$ directions) and breaks another set (in the $x_\perp$
directions).  However, a solution can be obtained that
preserves more translational symmetries by using
a more symmetric source, e.g., one supported on a line,
plane, or a higher dimensional surface.  Constructing such
a solution can be thought of as `smearing out' the charge
of a less symmetric solution.   Smearing out
a given brane solution often results in a spacetime with
a singular horizon.  However, this need not be especially
worrying if one regards the smeared solution as merely
an effective description analogous to describing a collection of
discrete atoms as a continuous fluid.  One imagines an array of branes in which
a large number of unsmeared basic
branes are placed in the spacetime with a small spacing between
the branes.

The next technique to discuss is that of combining the basic
types of charge.  As mentioned above, this is in general
rather difficult.  If, however, two solutions preserve some of the same
supersymmetries {\it and} they have been
engineered to have the same translation symmetries (for example, by
smearing), then they tend to be rather easy to combine.
Making a simple guess as to the way in which the
relevant harmonic functions ($H_2,H_5,H_{AS},H_M$) should enter the metric and
gauge fields tends to lead to a solution to the supergravity equations
which preserves the common supersymmetries.

So far as I know, there are no general theorems available on this subject.
We will thus content ourselves with a few simple examples which fall into the class discussed in \cite{Tharm,GKT}; see also \cite{45rev}.
We have already discussed the solution (\ref{M2}) corresponding to
a set of parallel M2-branes.  This solution preserves half of the
original 32 supersymmetries of 10+1 supergravity.  The particular
supersymmetries that are broken are related
to the plane in space along which
the M2-branes are oriented.   Let us call the spatial coordinates along
these branes $x^1_\parallel$ and $x^2_{\parallel}$.
We could also consider another set
of M2-branes oriented along another plane associated with two other
coordinates $y^1_{\parallel}$ and $y^2_{\parallel}$, which are to
be orthogonal to the $x_\parallel$ coordinates.  A set of  solutions containing
both types of branes and preserving the $8$ supersymmetries common to
both sets of M2-branes separately is given by

\begin{eqnarray}
\label{2M2}
A &=& H_x^{-1} dt \wedge dx_{\parallel,1} \wedge dx_{\parallel,2}
+ H_y^{-1} dt \wedge dy_{\parallel,1} \wedge dy_{\parallel,2}
\cr
ds^2 &=& - H_x^{-2/3} H_y^{-2/3} dt^2 +  H_x^{-2/3} H^{1/3}_y
 dx_{\parallel}^2 + H_x^{1/3}  H_y^{-2/3} dy_{\parallel}^2
 \cr &+& H_x^{1/3} H_y^{1/3} dx_{\perp}^2,
\end{eqnarray}
where $H_x$, $H_y$ are functions only of the six spatial coordinates $x_\perp$
that are transverse to both sets of branes.  The functions $H_x$ and
$H_y$ are, as usual, related to source distributions through
$\partial^2_{x_\perp} H_x =-7 \Omega_8 \rho_x$ and
$\partial^2_{x_\perp} H_y = - 7 \Omega_8 \rho_y$ and the distributions $\rho_x$
and $\rho_y$ may be arbitrary functions of $x_\perp$.

Note
that the form (\ref{2M2}) is just like that of (\ref{M2}) except that
we include two harmonic functions.  A given term in the metric (\ref{2M2})
is multiplied by a power of each harmonic function determined
by whether the term refers to distances along or transverse to the
corresponding brane.  These powers are identical to the ones in (\ref{M2}).

In the solution (\ref{2M2}), we
have taken the two sets of branes to be completely orthogonal to each other.
However, other choices of the relative angle still preserve the same
amount of supersymmetry.  If one thinks about the coordinates
$x_\parallel$, $y_\parallel$ as two sets of holomorphic coordinates on ${\bf C}^2$,
then the requirement for a supersymmetric solution is that
the $x_\parallel$ and $y_\parallel$ planes are related by a $U(2)$
transformation \cite{angles}
as opposed to a more general $O(4)$ transformation.
The metric in this case takes a similar form, with the part of the
metric on the four-space spanned by $x_\parallel$, $y_\parallel$
taking a certain Hermitian form.

However, combining the two sets of branes without first smearing them to
generate four translation symmetries is more difficult.
It turns out that, when one or both of the sets of branes is `localized' (i.e.,
not completely spread out along the other set of branes) then
the supergravity equations no longer cleanly divide into pieces
describing each set of branes separately.  The case where only
one set of M2-branes is localized (and thus two translational symmetries remain)
is still tractable, however.  The solution
still takes the same basic form (\ref{2M2}) and
construction of the solution
still splits into two parts.
One can first solve a standard flat space
Poisson equation for the harmonic function
$H_x$ associated with the delocalized set of branes. One then
has a linear differential equation to solve for the localized brane
harmonic function $H_y$, where
$H_x$ appears in the particular differential operator to be inverted.  Such solutions
exhibit an interesting phenomenon:  While localized solutions exist when the branes are separated (in the $x_\perp$ directions), the $y$-branes effectively delocalize along the $x$-branes in the limit where this separation is removed\footnote{This behavior is also typical of other BPS intersecting branes when the intersection (i.e., the set of directions common to both branes) is either 0+1 or 1+1 dimensional, though fully-localized solutions exist for higher-dimensional intersections. See \cite{SS,AP,pertdiv} for details and \cite{CHI,DEG1,DEG2} for examples of such fully localized solutions. }.

Let us now return to the smeared solution (\ref{2M2}) and consider
the case in which
$\rho_x = \rho_y = r_0^4 \delta^{(6)}(x_\perp)$.  We then find that $H_x$ and
$H_y$ diverge at $x_\perp = 0$ like $|x_\perp|^{-4}$.  As a result,
the 5-spheres at $x_\perp=0$ are infinite in volume and the
solution is somewhat singular.  However, adding a third M2-brane in
another completely orthogonal ($z_\parallel^1, z_\parallel^2$) plane
yields a non-singular solution.
The metric and gauge field

\begin{eqnarray}
\label{3M2}
A &=&  H_x^{-1} dt \wedge dx_{\parallel,1} \wedge dx_{\parallel,2}
+ H_y^{-1} dt \wedge dy_{\parallel,1} \wedge dy_{\parallel,2}
+ H_z^{-1} dt \wedge dz_{\parallel,1} \wedge dz_{\parallel,2}
\cr
ds^2 &=& - H_x^{-2/3} H_y^{-2/3} H_z^{-2/3}
 dt^2 +  H_x^{-2/3} H^{1/3}_y H_z^{1/3}
 dx_{\parallel}^2 + H_x^{1/3} H_y^{-2/3} H_z^{1/3} dy_{\parallel}^2 \cr
&+& H_x^{1/3} H_y^{1/3} H_z^{-2/3} dz_{\parallel}^2
 + H_x^{1/3} H_y^{1/3} H_z^{1/3} dx_{\perp}^2
\end{eqnarray}
for $\partial^2_{x_\perp} H_{x,y,z} (x_\perp) = -2 \Omega_3
\rho_{x,y,z} (x_\perp)$
yield a BPS solution
of the supergravity equations that preserves 1/8 of the supersymmetry
(i.e., 4 supercharges) and has a smooth horizon.  Moreover, this solution
has the property that the translational Killing fields
$\partial x^i_{\parallel}$,
$\partial y^i_{\parallel}$,
$\partial z^i_{\parallel}$ have norms that do not vanish on the horizon.

In contrast, recall that while the solution (\ref{M2}) for a single
M2-brane has a smooth horizon, the spatial translational Killing fields
have vanishing norm there.  As mentioned above, this means that compactifying
a single M2-brane by, for example, taking the coordinate $x_{\parallel}^1$
to live on a circle, yields a solution with a conical singularity at
the horizon and vanishing entropy.
On the other hand, because the norms of the spatial translations
do not vanish for the solution (\ref{3M2}), it compactifies nicely
into a black object with finite horizon area\footnote{As given by (\ref{3M2}) this horizon is in fact smooth. We expect the compactified black hole to be subject to an instability simmilar to that of extreme Reissner-Nordstrom so that the perturbed conformal diagram is given by Fig. 11.3. Recall, however, that this instability affects only the black hole interior and so does not change the horizon area.}.
This is the simplest BPS black brane solution with a finite entropy
and, as a result, it is the simplest solution for which
a microscopic accounting of the entropy has been given in string theory.
A straightforward calculation shows the the horizon area is
\begin{equation}
\label{3M2area}
A = \Omega_3 r_x r_y r_z L_{1x}L_{2x}L_{1y}L_{2y}L_{1z}L_{2z},
\end{equation}
where $\Omega_3$ is the volume of the unit three-sphere and
the $L$'s are the lengths of the various circles on which the solution
has been compactified.

Now, charges are quantized in string/M-
theory and it is useful to express the entropy in terms of the number
of charge quanta $Q_x,Q_y,Q_z$ carried by the various branes.
The tension of a single M2-brane is $(2\pi)^3l_p^{-3}$, where $l_p$ is the
eleven-dimensional Plank length, defined by
$16 \pi G_{11} = 2 \kappa_{11}^2 = (2\pi)^8 l_p^9$.
Note that $r^2_x$ is a measure
of the charge {\it density} of the $x$-type branes per unit cell of the $y,z$
four-space. As such, $r^2_x$ is proportional to $Q_x/L_{1y}L_{2y}L_{1z}L_{2z}$.
Inspection of the area formula (\ref{3M2area}) thus shows that
rewriting the area in terms of the integer charges will remove
the factors of $L$.  Putting in the proper normalization coefficients,
the result turns out to be
\begin{equation}
\label{BHE}
A/4G_{11} = 2\pi \sqrt{ Q_x Q_y Q_z}.
\end{equation}
We will comment briefly on the corresponding microscopic counting of
states in section \ref{SandD}.

\section{Kaluza-Klein and dimensional reduction}
\label{KKred}

So far, we have dealt almost exclusively with
eleven dimensional supergravity.  However, it is the 10-dimensional supergravity theories that
admit self-consistent
perturbative quantizations in terms of strings\footnote{It should be
mentioned that string theory is not a quantization of
pure 9+1 supergravity; string theory modifies the physics
even at the classical level, though only by adding heavy fields with masses of order $1/l_s$ where $l_s$ is the string scale.  See section \ref{bfe} for (a few) more
details.}.  This means
that the powerful technology of perturbative quantum field
theory can be brought to bear on questions concerning
quantum dynamics.  This perturbative technology can in particular
be applied to certain branes in 9+1 supergravity.
It is through this fusion of supergravity and
perturbative field theory that
string/M-theory has been revolutionized in recent years
via studies of duality, black hole entropy, and more recently
the Maldacena conjecture or AdS/CFT correspondence.

This article is not the place to enter into
a detailed discussion of string perturbation theory, though we will
comment briefly on the subject in section \ref{pert}.
The reader interested in learning that subject should
consult the standard references (e.g. \cite{GSW,Joe,CVJ,BBS,DineBook,Kbook}).
Our purpose here is to provide a clear
picture of the supergravity side of BPS brane physics, and in particular
to discuss their relationship with the eleven dimensional
theories.  As the so-called type IIA 10-dimensional theory can be obtained by Kaluza-Klein reduction from M-theory, we warm-up by discussing Kaluza-Klein
compactification in non-gravitational theories in section \ref{KK}.  We then address
the reduction of M-theory to the IIA theory in detail in section \ref{SUGRAVKK}, setting the stage for our discussion
of 9+1 branes in section \ref{10branes}.

\subsection{Some remarks on Kaluza-Klein reduction}

\label{KK}

The idea of the Kaluza-Klein mechanism is that, at low
energies,
a {\it quantum} field theory on an $n+d$-dimensional
spacetime in which $d$ of the dimensions are compact
behaves essentially like a quantum field theory on
an $n$-dimensional spacetime.  To see why, consider
a free scalar field on $M^n \times S^1$ where $M^n$ is
$n$-dimensional Minkowski space. Normal modes of the field are labeled by
an $n$-vector momentum $p$ and an integer $k$ corresponding
to momentum around the $S^1$.  Suppose that
the length of the $S^1$ is $L$,
so that the dispersion relation associated with one-particle
excitations is $E^2 = p^2 + (k/L)^2$.  If we now consider
the theory at energy scales less than $1/L$, the only
states with such a low energy have $k=0$; i.e., they are
translationally invariant around the $S^1$.

In this way, our scalar field reduces at low energies
to a quantum field on $n$-dimensional Minkowski space.
Note that this is an intrinsically quantum mechanical effect, associated with both the quantization of energy and with the discrete spectrum of the Laplacian on
a circle.  Since the Laplacian has a discrete spectrum on any
compact space, the same basic mechanism operates with any choice
of compact manifold.    The simplest cases to analyze
are those in which the spacetime is a direct product of
a non-compact spacetime $M$ with a compact manifold $K$, and
in which $K$ is a homogeneous space.
In that case, the lower dimensional (reduced) theory is typically obtained
from the higher dimensional one simply by taking the fields
to be invariant under the symmetry group that acts transitively on the compact
manifold.   On a general manifold of the form $M \times K$,
the reduced theory is given by considering the zero-modes
of the Laplacian (or other appropriate differential operator) on
$K$.  Similar, but less clean, mechanisms
may apply even when the spacetime is not a direct product of a compact
and a non-compact spacetime.

The effect of compactification on interacting fields is similar.
At the perturbative level the story is exactly the same, and
non-perturbative effects seldom change the picture significantly.

\subsection{Kaluza-Klein in (super)gravity}
\label{SUGRAVKK}

We now turn to Kaluza-Klein reduction in a theory with gravity.
Since the spacetime metric is dynamical, this case is perhaps
not as clean cut as the scalar field example just discussed.
However, at the perturbative level, one may treat
gravity just as any other field.  Our general experience with
quantum mechanics and the uncertainty principle also
makes it reasonable on more general grounds
to expect that excitations
associated with the small compact space will be expensive in
terms of energy.  Thus, at least at first glance, we expect
that gravity on a manifold of the form $M \times K$
reduces at low energies to a theory on the non-compact
manifold $M$.

We are most interested in Kaluza-Klein
reduction of eleven-dimensional supergravity on ${\cal M} = M \times S^1$ where $M$ is a 9+1 dimensional asymptotically flat spacetime.
We expect the reduced theory to be obtained by considering
the class of eleven-dimensional field configurations that
are translationally invariant around the $S^1$.  Let us therefore
assume that our eleven dimensional spacetime ${\cal M}$
has a spacelike Killing
vector field $\lambda^\mu$ whose orbits have the topology $S^1$.
It is convenient to normalize
this Killing field to have norm $+1$ at infinity
and to denote the length of the Killing orbits there by $L$.
The Killing field is not necessarily
hypersurface orthogonal.

Since the translation group generated by the
Killing field acts nicely (technically, `properly
discontinuously' \cite{HawkingEllis}) on our eleven dimensional
spacetime ${\cal M}$, we may consider
the quotient of the smooth topological space
${\cal M}$ by the action of this group.
The result is a new topological space $M$, which is a ten
dimensional smooth manifold.  This is the manifold on which our
$10=9+1$ dimensional reduced theory will live.

By using the metric,
we define a set of projection operations on the various 10+1
fields with each projection providing a different field in the 9+1
dimensional spacetime.
Recall that a field is an object which transforms in a certain
way under local Lorentz transformations (i.e., diffeomorphisms) of the
manifold.  The diffeomorphisms of the 9+1 manifold will be that
subgroup of the eleven-dimensional diffeomorphisms that leaves the
killing field $\lambda^\mu$ invariant.  Thus, the transformations
that become the diffeomorphisms of the $9+1$ manifold are a proper
subgroup of the 10+1 diffeomorphisms and a single 10+1 field
can contain several 9+1 fields.

To see how the 9+1 fields are constructed, consider any
coordinate patch $U$ (with coordinates $x^a$) on the 9+1 manifold $M$.
If $V \subset {\cal M}$ is the preimage of $U$ under
the above quotient construction, then each $x^a$ defines
a function on $V$.  Since no linear combination of the
gradients of the $x^a$ functions can be proportional to the Killing field
$\lambda^\mu$, we can complete this set of functions to
a coordinate patch on $V$ by adding a (periodic) coordinate $\theta$
which is proportional to the Killing parameter along any orbit of
$\lambda^\mu$; i.e., satisfying $\theta_{,\mu} \lambda^\mu
= \lambda^\mu \lambda_\mu$.

This coordinate system gives an explicit realization of the
natural decomposition of the 10+1 fields into a set of 9+1 fields.
The set of gradients $x^a_{,\mu}$ of the 9+1 coordinates
define a projection operation on any contravariant (upper) index, as does
the gradient $\theta_{,\mu}$ of the coordinate $\theta$.  Thus,
from the 10+1 contravariant metric $g^{\mu \nu}$, we can define the
9+1 metric $g^{ab} =  x^a_{,\mu} x^b_{,\nu} g^{\mu \nu}$,
a 9+1 abelian vector field $A_1^a = - x^a_{,\mu} \theta_{,\nu} g^{\mu \nu}$,
and a 9+1 scalar field $\phi$ through
$L e^{4\phi/3} = \lambda^\mu g_{\mu \nu} \lambda^\nu$.  The particular
coefficient of $\phi$ is chosen so that it is canonically
normalized\footnote{When $2\kappa_{10}^2$ is set to one, see below.}.
This $\phi$ is the famous dilaton of string theory, and it is this field
which is responsible for many of the differences between
supergravity in less than eleven dimensions and familiar
Einstein-Maxwell theory.
It is clear that all of these fields transform in an appropriate way under
9+1 diffeomorphisms.

We now make several important observations. The first is that nondegeneracy of the 10+1
metric implies non-degeneracy of the 9+1 metric.  Thus,
$g^{ab}$ has an inverse which gives the covariant metric $g_{ab}$.

The second is that the scalar has been defined by the
norm of the Killing field and not the norm of $\theta_{,\mu}$
as one might expect.  The point is that these two objects are related.
To see this, let us first note that the coordinates $x^a$
are constant along the orbits of the Killing field.
Thus, the Lie derivative of $x^a$ along $\lambda^\mu$ vanishes, and
we have $x^a_{,\nu} g^{\nu \mu} \lambda_{\mu} =0$.
This means that the gradients $x^a_{,\mu}$ span the space
orthogonal to $\lambda_\mu$ at each point.  But, by
definition, $\lambda^\mu \theta_{,\mu} = \lambda^\mu \lambda_\mu$.
Thus, we find that $\theta_{,\mu} - \lambda_{\mu}$ is of the
form $c_a x^a_{,\mu}$ where $c_a$ is some function on the 9+1
spacetime.  This
fact, together with
the definition of $A_1$, can be used to derive the relation:
\begin{equation}
c_a = - g_{ab} A_1^b.
\end{equation}
Thus, we have
\begin{equation}
\theta_{,\mu} \theta_,{}^\mu = \lambda^\mu \lambda_{\mu} + A_{1a}A_1^a.
\end{equation}
We see that the definition of
$\phi$ differs from the seemingly more natural one only
by a function of the vector field $A_1$.  Choosing to
write $\phi$ directly in terms of the Killing field $\lambda^\mu$
removes a mixing between the vector field and scalar that
would otherwise obscure the physics.
Note that we have related the scalar field $\phi$ to the logarithm
of the norm of the Killing field, and that this norm is
positive by assumption.

Finally, let us consider the vector field $A_1.$
Although
we have $x^a_{,\nu} g^{\nu \mu} \lambda_{\mu} =0$,
the vector field $A_1$ need not vanish.  Note that there is a freedom
to redefine the zero of $\theta$ at each value of the $x^a$.  This amounts
to the transformation $\theta \rightarrow \theta - \Lambda(x)$.  Under
this operation, we see that the $9+1$ metric $g^{ab}$ is not affected,
and neither is the scalar (since it depends only on the norm
of the Killing field)
while the vector field transforms as $A_1^a \rightarrow A_1^a +
\Lambda_{,b} g^{ab}$; i.e., $A_{1a} \rightarrow A_{1a} +
\Lambda_{,a}$.  Thus, we see that $A_1$ is in fact an abelian
gauge field.  The associated field strength $F_2=dA_1$ (with $A_1$ considered as a 1-form) is just the `twist' of the Killing field $\lambda^\mu$, which measures the the failure of $\lambda^\mu$ to be hypersurface orthogonal.

It is interesting to ask about the charge to which this gauge field
couples, as the field itself arose directly from the reduction of the
gravitational field in eleven dimensions.  To this end, consider a ``gauge" transformation of the above form with constant $\Lambda$.  In familiar Maxwell theory, this global U(1) rotation is generated by the total electric charge operator.  But in terms of ${\cal M}$ it is a shift of $\theta$, which is just a translation along $\lambda^\mu$.  We therefore identify the total $A_1$-charge on $M$ with the corresponding momentum on ${\cal M}$.   One may check that any timelike total energy-momentum for ${\cal M}$ becomes a charge and a ten-dimensional energy-momentum vector on $M$ satisfying a BPS bound.

In performing calculations, it is often useful to express
the above decomposition in terms of the eleven dimensional
{\it covariant} metric $ds_{11}^2$.  The reader may check that
we have
\begin{equation}
\label{ds11}
ds_{11}^2 = g_{ab} dx^a dx^b + e^{4 \phi/3}[d\theta + A_{1a} dx^a]^2.
\end{equation}

One might think that it is natural to decompose the antisymmetric 3-form $A_3$ into a 9+1 3-form $\hat A_{3}^{abc} = A_{3}^{\mu \nu \rho}
x^a_{,\mu}x^b_{,\nu} x^c_{,\rho}$ and a 2-form
$A_2^{ab} = A_{3}^{\mu \nu \rho}
x^a_{,\mu}x^b_{,\nu} \lambda_{\rho}$ in order that both be invariant under the gauge transformation $A_1 \rightarrow A_1 + d \Lambda_0$.  However, it turns out that the 3-form $\hat A_3$ then transforms non-trivially under the gauge transformations associated with the 2-form potential $A_2$.  The various gauge transformations cannot be completely disentangled and in fact the standard choice is instead to define $\tilde A_3, A_2$ by
\begin{equation}
A_3 = \frac{1}{3!}  \tilde{A}_{3 abc} dx^a\wedge dx^b \wedge dx^c
+ \frac{1}{2!} A_{2ab} dx^a \wedge dx^b \wedge d\theta.
\end{equation}
As a result, the 3-form $\tilde A_3$ is {\it not} invariant under $A_1 \rightarrow A_1 + d \Lambda_0$ but instead transforms as $A_3 \rightarrow A_2 \wedge d\Lambda_0$.  In addition,
the gauge symmetry
of the eleven dimensional $A_3$ implies that there are
9+1 gauge symmetries $A_3 \rightarrow A_3 + d \Lambda_2$
and $A_2 \rightarrow A_2 + d \Lambda_1$ where $\Lambda_n$
are arbitrary $n$-forms.
From here on, we drop the tilde ( $\tilde{}$ ) on $\tilde A_3$.
The decomposition of the fermionic
fields is similar, but we will not go into this in detail.

\subsection{On 9+1 dynamics: Here comes the dilaton} \index{dilaton}

The dynamics for the 9+1 theory follows from that of eleven dimensions
by inserting the
relations between the 9+1 fields and the 10+1 fields into the
action.  The result is an action principle for the
9+1 theory which takes the form

\begin{eqnarray}
\label{10daction}
S_{\rm 9+1,bosonic} &=& \frac{1}{2\kappa_{10}^2} \int d^{10} x
\Bigl[
\sqrt {-g} \left(e^{2\phi/3} R - \frac{1}{2} e^{2 \phi} |F_2|^2
\right) \cr
&-&
\frac{1}{4 \kappa^2_{10}} \int d^{10}x \sqrt{-g}
\left( e^{-2\phi/3}|F^2|_3 + e^{2\phi/3}
|\tilde F_4|^2 \right) \cr
&-& \frac{1}{4\kappa_{10}^2} \int A_2 \wedge F_4 \wedge F_4 \Bigr].
\end{eqnarray}
Here all quantities refer to the 9+1 dimensional fields and we have defined
$F_n = dA_{n-1}$ and $\tilde F_4 = d A_3 - A_1 \wedge F_3$.  As opposed to $F_4$ itself, the new field strength $\tilde F_4$ is invariant under gauge transformations of the $A_1$ potential.
We have also defined $\kappa_{10}^2 = \kappa_{11}^2/L$.

An important feature of (\ref{10daction}) is that
the field $\phi$ appears all over the place, with
different factors of $e^\phi$ appearing in different terms.
The upshot of this is that the various gauge fields do not
couple minimally to the metric $g$.  Of course, we have the freedom to
mix the metric with $\phi$ by rescaling the metric
by some power of $e^\phi$.  This can be used to
make any one of the gauge fields couple minimally to the
new metric, or to remove the factors of $e^\phi$ in front
of the scalar curvature term and put the action in a form
more like that of familiar Einstein-Hilbert gravity.  However,
because of the way that different factors of $e^\phi$ appear
in the different terms, this cannot be done for all fields
at once.  Thus, we may think of each different gauge field
as coupling to a different metric.

A short calculation shows that the gauge fields $F_2$
and $F_4$ couple minimally to $e^{2\phi/3}g$ while the gauge
field $F_3$ couples minimally to $e^{-\phi/3}g$.
In doing this calculation, it is important to realize that
terms like $|F_2|^2$ contain implicit factors of the metric $g$ (see \ref{Fnorm})
which has been used to contract the indices.
On the other
hand, it is for the `Einstein metric' $e^{\phi/6}g$
that the gravitational part of the action takes the standard
Einstein-Hilbert form (the integral of the scalar curvature
density) without any extra factors of $e^\phi$.

The choice
of a particular metric in the class $e^{\alpha \phi} g$
is known as the choice of conformal frame.  One can make a choice
of frame that simplifies a given calculation, if one desires.
It is interesting to note that in the
conformal frame defined by (\ref{ds11}) the field $\phi$ has no explicit
kinetic term so that its variation leads to a constraint.
It turns out that this is just a combination of the usual
constraints that one would expect in a gravitating theory,
and that a term of the form $\partial_a \partial^a \phi$ does
appear in the equations of motion obtained by varying the metric
in that frame.

The two most useful choices of conformal frame are the
Einstein frame (defined by the Einstein metric $g_E = e^{\phi/6}g$ discussed above)  and the so-called string frame.
The action in the Einstein frame is a handy thing to have
on hand, so we will write it down here.   If we
now let $g_E$ denote the metric in the Einstein frame and let
$R_E$ be the associated curvature, the
action is
\begin{eqnarray}
\label{Eaction}
S_{\rm IIA, bosonic} &=& \frac{1}{2\kappa_{10}^2} \int d^{10}x
\sqrt {-g_E} \left(R_E - \frac{1}{2} \partial_a\phi \partial^a \phi\right)
\cr &-&
\frac{1}{4 \kappa_{10}^2} \int d^{10}x
\sqrt{-g_E} \left( e^{3 \phi/2} |F_2|^2 + e^{-\phi} |F^2|_3 + e^{\phi/2}
|\tilde F_4|^2 \right) \cr
&-& \frac{1}{4 \kappa_{10}^2} \int  A_2 \wedge F_4 \wedge F_4 .
\end{eqnarray}
Note that,
in Einstein frame, the gauge fields are all sources for the
dilaton but the metric is not.  Also, since the kinetic term for the
dilaton now takes the standard form, we can see that the dilaton would
be canonically normalized if we set $2 \kappa_{10}^2$ to one.
Finally, since it is in this frame that the gravitational dynamics takes
the familiar Einstein-Hilbert form, this is the frame in which the
standard ADM formulas for energy and momentum may be applied and in which the entropy of black holes is given by $A/4$ in Planck units.

The string frame \index{string frame} is defined by taking the metric to be
$e^{2\phi/3} g$, where $g$ is the original metric that appeared in (\ref{ds11}).  Thus the string metric $g_S$ and the Einstein
metric $g_E$ are related by  $ds_{E} = e^{-\phi/2} ds_{S}$ and the action in the string frame takes the form
\begin{eqnarray}
\label{Stringaction}
S_{\rm IIA, bosonic} &=& \frac{1}{2\kappa_{10}^2} \int d^{10}x
\sqrt {-g_S} e^{-2\phi}
\left(R_S + 4 \partial_a\phi \partial^a \phi - \frac{1}{2}|F_3|^2
\right) \cr &-&
\frac{1}{4 \kappa_{10}^2} \int d^{10}x
\sqrt{-g_S} \left( |F_2|^2 + |\tilde F_4^2| \right) \cr
&-& \frac{1}{4 \kappa_{10}^2} \int  A_2 \wedge F_4 \wedge F_4 .
\end{eqnarray}
After setting $c=\hbar =1$, the parameter $\kappa^2_{10}$ has units
of $(length)^8$.  It is useful to write $2 \kappa_{10}^2 = (2 \pi)^7
g_s^2 l_s^8$ where $l_s$ is the ``string length'' \index{string length} and $g_s$ is the
``string coupling.''    \index{string coupling} For more on the separate role of
$g_s$ and $l_s$, see section \ref{pert}.

Note that two of the gauge fields ($F_2$ and $F_4$) couple minimally
to the string metric $g_S$.  These two gauge fields are known as
{\it Ramond-Ramond} (R-R) gauge fields \index{Ramond-Ramond gauge fields} while $F_3$ is known as
the {\it Neveu-Schwarz Neveu-Schwarz} (NS-NS) gauge field, or sometimes just as the Neveu-Schwarz (NS)  gauge field \index{Neveu-Schwarz gauge field} for short\footnote{In a confusing piece of terminology, it is sometimes also called the Kalb-Ramond gauge field.}.
For an explanation of how this terminology arose in string
perturbation theory, see e.g. \cite{Joe,CVJ,BBS,Kbook}.  The potential
$A_{2}$ for this field is commonly written $B_2$ (and its field strength $F_3$ is written $H_3$) and when
string theorists discuss ``the B-field,'' it is this potential to which
they are referring.

What makes the string metric especially useful
is that it turns out to be the metric to which fundamental
strings (which we have not yet discussed) couple minimally, and thus in which one makes the
most direct contact with string perturbation theory.
This, however, is a discussion for another place and time.

In the above, we have discussed only the compactification of eleven dimensional
supergravity on a circle.  One can, of course, consider further
compactifications to smaller dimensional manifolds.  The story in that
case is much the same except that the number of lower-dimensional
fields generated increases rapidly.  In particular, further compactification
generates large numbers of massless scalars that couple non-minimally
to the various gauge fields.  These cousins of the dilaton are generally
referred to as {\it moduli}. \index{moduli}

All of these moduli have a tendency
to diverge at the horizon of an extreme black hole, making the solution
singular.  One may think of the issue as follows:  the moduli, like
the dilaton, couple to the gauge fields so that the squared
field strengths $F^2$ act
as sources.  This can be seen from the action (\ref{Eaction}) in the
Einstein frame.  Non-singular extremal black hole solutions typically have
an infinite throat, as in the four and five dimensional Einstein-Maxwell
examples discussed earlier.  This means that a smooth such solution
would have an infinite volume of space near the horizon in which
the gauge field strengths are approximately constant.  Unless
these gauge fields are tuned to have $F^2 =0$ or the various
gauge fields are somehow played off against one another, this provides an infinite source for the moduli.  As a result, smooth solutions are obtained only when the charges of the black hole are such that the potential for the moduli provided by the various $F^2$ terms has a stationary point.  In the spacetime solution, the moduli then approach this stationary point as one approaches the horizon.
This phenomenon is known as the attractor mechanism \index{attractor mechanism} for extreme black holes, see e.g. \cite{attract1,attract2}.
As a result of this effect, some care is required to construct an
extremal black hole solution with a smooth horizon and such
solutions necessarily carry more than one charge.
For a brane solution, the norm of each spacelike Killing field acts like
a modulus whose sources must be properly tuned (as the norm would define a new dilaton-like scalar under further Kaluza-Klein reduction).

This is essentially
the issue encountered at the end of section \ref{smear} in which
it was found that three charges (in the case, three different types of
M2-branes) were required to obtain a brane solution in which the norms
of the spacelike Killing fields did not vanish on the horizon.  Recall
that a Killing field with positive norm allows us to Kaluza-Klein
reduce the spacetime to a solution of lower-dimensional supergravity.
Because the three charge solution (\ref{3M2}) has six
Killing vector fields whose norms do not vanish on the horizon, it
may be reduced all the way down to a solution of 4+1 gravity.
In this context, it represents an extreme black {\it hole}.
In fact, it reduces to just the standard 5+1 extremal black hole
(\ref{5RN}) of Einstein-Maxwell theory.

By the way, the theory discussed above
is far from the only supergravity theory in ten
dimensions.  It is a particular kind called `type IIA,' originally constructed in \cite{CWIIA,HNIIA,GPIIA}.  The `II' refers
to the fact that there are two independent gravitino fields and, as a result, two 10-dimensional spinors  worth of supercharges.  Each 10-dimensional spinor has 16 components, so this theory is maximally symmetric just like the 11-dimensional theory.  In type IIA theory, these
gravitinos have opposite chirality.  This is turn allows type IIA
theory to be defined even on non-orientable manifolds.  There
is also a type IIB theory which has two gravitinos (and is thus also maximally supersymmetric), but of the
same chirality.  Thus, type IIB theory can only be defined on manifolds
with a global notion of chirality and, in particular, only on
orientable spacetimes.  We will discuss type IIB theory further in
section \ref{T} below.  Two other supergravity theories with less
supersymmetry (only 16 supercharges) are known as the type I and heterotic theories.  Each of these types of supergravity in ten dimensions is associated with its own
version of string theory.  We will not discuss type I or heterotic
supergravity here, but a discussion of these theories and how they
are related to the type II theories can be found in \cite{Joe,CVJ,BBS,DineBook,Kbook}.

\section{Branes in 9+1 type II supergravity}
\label{10branes}
\label{branesols}

We now wish to discuss the basic brane solutions of type II supergravity
in 9+1 dimensions.  Since any solution of type IIA theory is really
a solution of eleven-dimensional supergravity in disguise, any brane solution of
type IIA theory immediately defines a brane solution of eleven
dimensional supergravity.  Thus, we should be able to construct the basic
brane solutions of type IIA theory by working with the basic brane
solutions of section \ref{branes}.    For this reason we address
the type IIA solutions first in section \ref{IIAbranes}.
Next follows a short aside on brane singularities in section \ref{bsing}.
We then briefly discuss
type IIB supergravity, and its relation through so-called
T-duality with the type IIA theory, in section \ref{T}. Below, we will discuss only
branes which become BPS in the extreme limit, although intrinsically non-BPS D-branes can also be of interest (see e.g.
\cite{non-BPS,non-BPSII}).

\subsection{The type IIA branes}
\label{IIAbranes}

In our decomposition of the eleven-dimensional
metric and gauge field into the various fields of ten dimensional
supergravity, we proceeded by projecting the fields along directions both parallel and transverse to the Kaluza-Klein Killing field $\lambda^\mu$.  In order to get
brane solutions of type IIA theory that are charged under all of the
type IIA gauge fields, a similar operation will need to be performed on
the 11-dimensional branes.  For any given brane in eleven dimensions, we
will need to reduce both a basic brane solution in which the Killing
field acts along the brane (i.e., is a symmetry of the brane), and one in
which it acts transverse to the basic brane.

One may at first
wonder what it means for the brane to be transverse to the Killing field
since translations along a Killing field must leave the solution invariant,
and therefore must preserve the brane.  The answer to this puzzle
is the smearing mentioned in section \ref{smear}.  One can take a basic
brane solution, pick a direction transverse to the brane, smear
the brane in that direction, and then reduce to 9+1 dimensions along the smearing direction.

Performing the required reductions amounts to no more than using the
relations between the 9+1 fields and the 10+1
fields given in section \ref{KKred}
to write down the 9+1 solutions from the branes given
in section
\ref{branes}.  We leave the details of the calculations to the reader,
but we provide a list here of the various 9+1 brane solutions.  Below,
we group together those branes charged under the Ramond-Ramond
gauge fields and those charged under the NS-NS gauge fields.  This
grouping is natural from the
point of view of the type IIA theory (and of string perturbation theory),
though we will see that it is somewhat less natural from the eleven
dimensional point of view.

Let us begin with the Ramond-Ramond branes.   It turns out that
the type IIA theory has $p$-brane solutions with Ramond-Ramond charge
for every even $p$.  What is very nice is that, in terms of the
string metric, all of these solutions take much the same simple
form.   In order to treat all of the branes at once, it is useful
to introduce a uniform notation for both electrically and magnetically charged
branes.  For each
gauge field $A_n$, we can introduce (at least locally) a magnetic
dual gauge field $A_{9-n}$ through\footnote{Here we again ignore the Chern-Simons terms, which modify and complicate this simple uniform expression.  As usual, this suffices for the solutions discussed below.}
$dA_{8-n} = \star F_{n+1}$.  A brane which couples magnetically to $A_n$
then couples electrically to $A_{9-n}$ and vice versa.  In type IIA
theory, this notation should introduce no confusion as the
standard gauge fields have $n=1,2,3$ while these new (dual) gauge
fields have $n=5, 6, 7$.

Introducing the usual set of $p+1$ coordinates $x_\parallel$
along the brane and $9-p$ coordinates $x_\perp$ transverse to the brane
we have, for all even $p$,
\begin{eqnarray}
\label{Dsol}
ds_{string}^2 &=& H_p^{-1/2} dx_{\parallel}^2 + H_p^{1/2} dx_{\perp}^2
\cr
A_{p+1} &=&  H_p^{-1} dx_{\parallel}^0 \wedge ... \wedge dx_{\parallel}^p \cr
e^{2 \phi} &=& H_p^{(3-p)/2},
\end{eqnarray}
where $H_p$ is a function only of the $x_\perp$ coordinates
and satisfies
\begin{equation}
\label{Dlap}
\partial_\perp^2 H_p =  - (7-p) \Omega_{8-p} r_0^{7-p} \delta^{(9-p)}(x_\perp)
\end{equation}
for the basic brane solution.     Here $\Omega_{8-p}$ is the volume
of the unit $(8-p)$-sphere and $r_0$ is a length scale parametrizing the strength of the source. These are the solutions known as extreme R-R $p$-branes or, in a slight \index{D-brane}
abuse of language, as (extreme) D$p$-branes, where the notation ``D" comes from the way these objects are described in string perturbation theory (where they are associated with Dirichlet boundary conditions for strings, see section \ref{SandD}).
As usual, we also obtain a
solution by considering more general source terms on the RHS of eq. (\ref{Dlap}).
For odd $p$, there are no gauge fields $A_{p+1}$ in type IIA supergravity
so (\ref{Dsol}) does not yield a solution to this theory for such
cases\footnote{One might also ask about the case $p=8$, since we
have not discussed a 9-form gauge potential.  It turns out that
there is in fact a Ramond-Ramond 8-brane in type IIA theory and that
its existence is tied to the Chern-Simons term in the type IIA action.
In this work, we follow a policy of considering only the asymptotically
flat brane solutions, which restricts us to branes of co-dimension 3 or higher; i.e., to $p \le 6$ in 10-dimensions.}.

Let us briefly mention that the non-extreme R-R brane solutions (for $p \le 6)$ take the rather simple form \cite{HSbs}
\begin{eqnarray}
\label{NEDsol}
ds_{string}^2 &=& H_p^{-1/2} \left(-f dt^2 + \sum_{i=1}^p (dx_{\parallel}^i)^2 \right)  + H_p^{1/2} \left(\frac{dr^2}{f} + r^2 d \Omega_{8-p}^2 \right)
\cr
A_{p+1} &=&[1+  \coth \beta (H_p^{-1}-1)] dx_{\parallel}^0 \wedge ... \wedge dx_{\parallel}^p \cr
e^{2 \phi} &=& H_p^{(3-p)/2},
\end{eqnarray}
where
\begin{equation}
 H_p = 1 + \frac{\sinh^2 \beta \ r_+^{7-p}}{r^{7-p}}, \qquad f = 1- \frac{r_+^{7-p}}{r^{7-p}}
 \end{equation}
  and $r_+, \beta$ specify the charge $Q$, mass $M$ per unit $p$-volume $V_p$, temperature $T$, and entropy density $S/V_p$ through
\begin{eqnarray}
Q &=& \frac{(7-p)\Omega_{8-p}}{2\kappa^2_{10} } r_+^{7-p} \sinh \beta \cosh \beta \cr
M/V_p &=& \frac{(8-p) \Omega_{8-p} r_+^{7-p}}{2\kappa_{10}^2 } \left(1 + \frac{7-p}{8-p}\sinh^2 \beta \right) \cr
T &=& \frac{7-p}{4\pi r_+ \cosh \beta} \cr
S/V_p &=& \frac{4\pi \Omega_{8-p}}{2 \kappa^2} \cosh \beta \ r_+^{(8-p)}.
\end{eqnarray}
In particular, the extremal limit is $\beta \rightarrow \infty$, $r_+ \rightarrow 0$ with $M/V_p$ fixed so that $M/V_p \rightarrow Q$ and $S \rightarrow 0$. For $p<5$, the temperature also vanishes in the extremal limit. However for $p=5$, $T$ remains nonzero, and for $p=6$ it diverges.
Here we allow only a single brane (analogous to having a single delta-function source in \ref{Dlap}) as non-extremal branes attract each other and solutions with more than one such brane are not stationary.   As for the M-branes, their global structure of (\ref{NEDsol}) is like
that of the Schwarzschild solution as opposed to that of
non-extremal Reissner-Nordstr\"om.

Although all of these R-R branes take the same simple form (\ref{Dsol}),
they proceed by quite different routes from the
eleven dimensional branes.  A short list follows:  The D0-brane
solution follows by reducing the smeared Aichelberg-Sexl metric  along the smearing
direction.
The D2-brane follows by reducing the smeared M2-brane along the smearing
direction.
The D4-brane is the reduction of the unsmeared M5-brane in a direction along
the brane. Finally, the D6-brane is the reduction of the unsmeared Kaluza-Klein
monopole along the $S^1$ fibers.

Next, there are the Neveu-Schwarz branes.  Since the only Neveu-Schwarz
gauge field is $A_2$, we expect to find two types of Neveu-Schwarz
branes.  The gauge field $A_2$ should couple electrically to a 1-brane
(a string) and it should couple magnetically to a 5-brane.  The
1-brane follows by reducing the M2-brane in a direction along the
brane.  The resulting solution
\begin{eqnarray}
\label{Fstring}
ds^2_{string} &=&  H_F^{-1} dx_\parallel^2 + dx_\perp^2 \cr
A_{2} &=& H_F^{-1} dx^0_\parallel \wedge dx^1_\parallel
\cr
e^{2\phi} &=& H_F^{-1}
\end{eqnarray}
is known as the {\it fundamental string}. \index{fundamental string} The reason for this is
that this solution represents the classical limit of a long, straight
version of the same string that appears in string perturbation
theory.

The Neveu-Schwarz 5-brane (NS5-brane) \index{Neveu-Schwarz 5-brane} is constructed by smearing
the M5-brane in a transverse direction and then reducing along the
smearing direction.  The result is
\begin{eqnarray}
\label{NS5}
ds^2_{string} &=&  dx_\parallel^2 + H_5 dx_\perp^2 \cr
F_{3} &=& - \frac{1}{3!} \partial_{x_\perp^i} H_5 \epsilon_{ijkl}
 dx^j_\perp \wedge dx^k_\perp \wedge dx^l_\perp
\cr
e^{2\phi} &=& H_5.
\end{eqnarray}

An interesting property of the NS5-brane is that, in the string metric,
the timelike Killing field has no horizon; its norm is constant across
the spacetime.  The would-be horizon at $x_\perp =0$ has receded
to infinite proper distance in all directions, not just along a Killing
slice as for the extreme Reissner-Nordstr\"om black hole.  As a result,
the coordinate patch above actually covers a manifold that, in the
string frame, is geodesically complete\footnote{However, it not
geodesically complete either in the Einstein frame or as viewed from
the eleven-dimensional perspective.  In each of these cases, there
is a null singularity at the horizon.}.

Finally, there are the purely gravitational `branes' given by
the 9+1 versions of the Aichelburg-Sexl metric and of the Kaluza-Klein
monopole with all gauge fields set to zero and constant dilaton.  These
may be either written down directly by analogy with (\ref{Mwave})
and (\ref{11KK}) or constructed by reducing the 10+1 solutions to 9+1 dimensions (and first smearing the Aichelberg-Sexl metric in some $x_\perp$ direction).

This exhausts the possible
ways to make extremal
9+1 branes by reducing (and perhaps smearing once) the basic
eleven-dimensional branes.    Below, we provide a few
words on their global structure and singularities.

\subsection{On brane singularities}
\label{bsing}

We have constructed the D-brane spacetimes from what (in most cases)
are smooth eleven-dimensional solutions (at least outside some horizon).  However,
the reduced solutions contain new singularities.  Of the
9+1 branes, only the NS 5-brane does not have a naked
singularity\footnote{This statement refers to the metric in the string
frame.  In the Einstein frame there is a naked singularity on the horizon.
Its story is much like that of the D2-brane discussed below.}.
For the D4- and D6-branes and the fundamental string,
this happens because the Killing field used in the reduction
has fixed points so that $\lambda^\mu\lambda_\mu$  and thus $e^\phi$ vanishes (and $\phi \rightarrow -\infty$).

For the D6-brane, the (6+1)-plane of fixed points
at $x_\perp=0$ is manifest in (\ref{11KK}) and results in a 6+1 dimensional timelike singularity in the 9+1
dimensional D6-brane solution.  The conformal diagram for the
D6-brane is therefore the one given in Fig. 11.5 below.

%
%  Created by WinFIG version 4.71
%  METADATA <version>1.0</version>
%

\vbox{
\centerline{
\begin{picture}(999,1865)(5314,-2473)
%  METADATA <id>2</id>
\thinlines
{\color[rgb]{1,0,0}\multiput(5401,-2461)(-3.75000,7.50000){21}{\makebox(1.6667,11.6667){\tiny.}}
\multiput(5326,-2311)(3.75000,7.50000){21}{\makebox(1.6667,11.6667){\tiny.}}
\multiput(5401,-2161)(3.75000,7.50000){21}{\makebox(1.6667,11.6667){\tiny.}}
\multiput(5476,-2011)(-3.75000,7.50000){41}{\makebox(1.6667,11.6667){\tiny.}}
\multiput(5326,-1711)(3.75000,7.50000){41}{\makebox(1.6667,11.6667){\tiny.}}
\multiput(5476,-1411)(-3.12500,7.81250){49}{\makebox(1.6667,11.6667){\tiny.}}
\multiput(5326,-1036)(4.68750,7.03125){33}{\makebox(1.6667,11.6667){\tiny.}}
\multiput(5476,-811)(-3.75000,7.50000){21}{\makebox(1.6667,11.6667){\tiny.}}
}%
%  METADATA <id>3</id>
\put(6001,-2236){\makebox(0,0)[lb]{\smash{{\SetFigFont{17}{20.4}{\rmdefault}{\mddefault}{\updefault}${\cal I}^-$}}}}
%  METADATA <id>4</id>
\put(6001,-1036){\makebox(0,0)[lb]{\smash{{\SetFigFont{17}{20.4}{\rmdefault}{\mddefault}{\updefault}${\cal I}^+$}}}}
%  METADATA <id>1</id>
{\color[rgb]{0,0,0}\put(5401,-661){\line( 1,-1){900}}
\put(6301,-1561){\line(-1,-1){900}}
}%
\end{picture}%
}
\centerline{Figure 11.5 Conformal diagram for the extreme D6-brane.}
}

\medskip
\noindent
Turning to the D4 brane, the fixed points of the Kaluza-Klein Killing field in the compactified M5-brane solution are less obvious from (\ref{M5}) but were briefly discussed in section \ref{branes}. There we saw that the fixed point set consisted of certain horizon generators and so defined a null plane.  As a result, the singularity of the D4-brane is null and lies on the would-be horizon.  \index{singularity!null} Because $\phi \rightarrow -\infty$, in terms of the 9+1 metric this turns out to be a (null) curvature singularity.   The story of
the fundamental string is much the same.  See Fig. 11.6 for both.

\vbox{\centerline{
\begin{picture}(2108,1936)(3731,-2473)
%  METADATA <id>1</id>
\thinlines
{\color[rgb]{1,0,0}\multiput(4651,-661)(3.75000,-7.50000){21}{\makebox(1.6667,11.6667){\tiny.}}
\multiput(4726,-811)(7.50000,-3.75000){21}{\makebox(1.6667,11.6667){\tiny.}}
\multiput(4876,-886)(4.68750,-7.03125){33}{\makebox(1.6667,11.6667){\tiny.}}
\multiput(5026,-1111)(7.50000,-3.75000){21}{\makebox(1.6667,11.6667){\tiny.}}
\multiput(5176,-1186)(4.68750,-7.03125){33}{\makebox(1.6667,11.6667){\tiny.}}
\put(5326,-1411){\line( 3,-2){225}}
}%
%  METADATA <id>2</id>
{\color[rgb]{1,0,0}\multiput(5537,-1561)(-7.50000,-3.75000){21}{\makebox(1.6667,11.6667){\tiny.}}
\multiput(5387,-1636)(-3.75000,-7.50000){21}{\makebox(1.6667,11.6667){\tiny.}}
\put(5312,-1786){\line(-3,-2){225}}
\multiput(5087,-1936)(-3.75000,-7.50000){21}{\makebox(1.6667,11.6667){\tiny.}}
\put(5012,-2086){\line(-3,-2){225}}
\multiput(4787,-2236)(-4.68750,-7.03125){33}{\makebox(1.6667,11.6667){\tiny.}}
}%
%  METADATA <id>3</id>
{\color[rgb]{0,0,0}\put(4643,-661){\line(-1,-1){900}}
\put(3743,-1561){\line( 1,-1){900}}
}%
%  METADATA <id>4</id>
\put(3899,-2326){\makebox(0,0)[lb]{\smash{{\SetFigFont{17}{20.4}{\rmdefault}{\mddefault}{\updefault}${\cal I}^-$}}}}
%  METADATA <id>5</id>
\put(3993,-982){\makebox(0,0)[lb]{\smash{{\SetFigFont{17}{20.4}{\rmdefault}{\mddefault}{\updefault}${\cal I}^+$}}}}
%  METADATA <id>8</id>
\put(5471,-1863){\makebox(0,0)[lb]{\smash{{\SetFigFont{17}{20.4}{\rmdefault}{\mddefault}{\updefault}singularity}}}}
%  METADATA <id>9</id>
\put(5824,-1563){\makebox(0,0)[lb]{\smash{{\SetFigFont{17}{20.4}{\rmdefault}{\mddefault}{\updefault}null}}}}
\end{picture}%
} \centerline{Figure 11.6 Conformal diagram
for the extreme D4-brane or fundamental string.}
}

\medskip

Let us now consider the D2-brane solution, which is the reduction of a
smeared M2-brane.
Although smearing the M2-brane in a transverse direction makes the
horizon of the eleven dimensional solution
 singular, one may take the perspective that the smeared solution
represents the approximate solution for an array of M2-branes
for which the length scale associated with the charge of each brane is much larger than the
spacing between the branes.  In this case, one interprets
the D2-brane horizon as being non-singular\footnote{Actually,
as consisting of many separate non-singular horizons.}
from the eleven dimensional point of view.  However,
from the 9+1 perspective, we clearly have $\phi \rightarrow + \infty$ on the horizon.  The metric (\ref{Dsol}) with $p=2$ also has a null curvature singularity on the horizon, though all curvature scalars are finite. The simplest way to detect the singularity from
the metric (\ref{Dsol}) is to note that the spheres around the brane
shrink to zero size at the horizon so that, if the solution were
smooth, the horizon could have only a single null generator, which is
impossible.
Thus the conformal diagram of the 9+1 D2-brane
solution is again given by Fig. 11.6.

With an eleven dimensional perspective in mind, the singularities
of the D2-, D4-, and D6-brane solutions might not be considered
especially troubling.
Nevertheless, from the 9+1 perspective
the singularities are quite real and represent places where the
9+1 equations of motion break down.   Let us recall that the Aichelberg-Sexl metric (the lift of the D0-brane solution to eleven dimensions)
{\it is} singular and should be thought of as describing
the approximate field produced by some `source.'  For the Aichelberg-Sexl metric,
one may think of this source as being a short wavelength
graviton, with the solution (\ref{Mwave}) itself representing just
the Coulomb part of the field.  Similarly,
looking at the way that the D-brane
singularities interact with the equations of motion through
(\ref{Dlap}), it is natural to think of the singularities as
representing bits of matter, like a braney form of extremal dust,
which are coupled to the supergravity.    By the way, this dual perspective of thinking of branes either as solitonic objects intrinsic to some basic version of the theory
(like supergravity or string theory) or as external objects or
sources coupled to such a theory is pervasive in string/M-theory.

\subsection{The type IIB theory and S- and T-dualities}
\label{T}

The other type of maximally supersymmetric gravity theory in 9+1
dimensions is called type IIB theory (originally constructed in \cite{SWIIB,HWIIB,JSIIB}).  It is not given by the dimensional
reduction of a 10+1 theory, though it has many of the same properties
as the type IIA theory.  For example,  both theories consist of a metric, a dilaton, and a $B_2$-field (which together form the so-called Neveu-Schwarz sector of the theory) together with various Ramond-Ramond gauge fields.  The theories are identical in the Neveu-Schwarz sectors,
so that the Aichelberg-Sexl, fundamental string, NS5-brane, and Kaluza-Klein monopole solutions are the same in both cases.
The main difference is that while IIA supergravity
has (electric and magnetic) Ramond-Ramond gauge fields $A_p$ of every odd rank, the type
IIB theory has Ramond-Ramond gauge fields $A_p$ of every even rank.  As in IIA, the Ramond-Ramond fields are minimally-coupled to the string-frame metric.  As a result, the D$p$-brane solutions in IIB are again given by (\ref{Dsol}), though now $p$ is odd; i.e., the IIB theory has D1, D3, D5, D7, and D9-branes\footnote{If one is interested in Euclidean solutions then it also makes sense to consider D(-1)-branes (D-instantons) in the IIB theory.}.

The D3-brane has particularly noteworthy features.  Since the Ramond-Ramond fields are minimally coupled in the string frame, the kinetic term of $A_4$ is of the form $\sqrt{-g_S} |F_5|^2.$  Note that this term is conformally invariant in 10-dimensions (just as the usual Maxwell kinetic term $\sqrt{-g} |F_2|^2$ is conformally invariant in 4-dimensions).  As a result, $A_4$ remains minimally coupled in any conformal frame, and in particular in the Einstein frame where the kinetic terms of the graviton and the dilaton have been diagonalized.  Thus, in contrast to all other IIB gauge fields, the 5-form $F_5$ does not act as a source for the dilaton!  Indeed, consulting (\ref{Dsol}) one finds that $\phi =  constant$ for $p=3$.  As a result the D3-brane has a smooth horizon.  In fact, the solution very similar to that of the M5-brane, having the same singularity-free conformal diagram (Fig. 11.4). The near-horizon geometry is AdS${}_5 \times S^5$.

Clearly the D3-brane is the marginal case that separates D$p$-branes with $p < 3$ from those with $p > 3$.  As one might expect, the D1 (and D(-1)) solution resembles that of the IIA D0,D2 solutions (with a null singularity where $\phi \rightarrow +\infty$) while the D5 resembles the D4 (with a null singularity where the Ricci scalar diverges).  Branes with large $p=7,9$ are special cases which we ignore here as they are not asymptotically flat.

To fully understand the relations between the various brane solutions it is important to understand certain symmetries known as S- and T-duality.  We begin with S-duality, which is a symmetry of IIB supergravity alone.  \index{S-duality} To study this symmetry, it is useful to define a complex scalar $\tau = A_0 + i e^{-\phi}$, a $2\times 2$ matrix ${\cal M}_{ij} = \frac{1}{{\rm Im} \ \tau} \left[{{|\tau|^2}\atop {- {\rm Re}\ \tau }}{{- {\rm Re}\ \tau }\atop 1} \right]$, and a vector of 2-form potentials $A_2^i = \left[{{B_2} \atop {A_2}} \right]$,
where we have called the Neveu-Schwarz gauge field $B_2$ (with field strength $H_3$) to distinguish it from the Ramond-Ramond gauge field $A_2$ (with field strength $F_3$).  In terms of these fields, the Einstein frame action\footnote{The equations of motion that follow from this action must be supplemented by the constraint that the 5-form $F_5 =  d A_4$ is self-dual; i.e., that $\star F_5=F_5$.} takes the form \cite{BBS}
\begin{eqnarray}
\label{IIBStringaction}
S_{\rm IIB, bosonic} &=& \frac{1}{2\kappa_{10}^2} \int d^{10}x
\sqrt {-g_E}
\left(R_E - \frac{1}{12} F^i_{abc} {\cal M}_{ij} F^{jabc} + \frac{1}{4}  (\partial^a {\cal M}_{ij} \partial_a {\cal M}^{-1ij})
\right) \cr &-&
\frac{1}{8 \kappa_{10}^2} \int d^{10}x
\sqrt{-g_S} |\tilde F_5|^2
- \frac{1}{4 \kappa_{10}^2} \int   A_4 \wedge  H_3 \wedge F_3,
\end{eqnarray}
where $F^i_3 = \frac{1}{3!} F^i_{abc} dx^a \wedge dx^b \wedge dx^c = dA^i_2$ and
$\tilde F_5 = F_5 + \frac{1}{2} \epsilon_{ij} A_2^i \wedge F_3^j$ with $\epsilon_{ij}$ antisymmetric and $\epsilon_{12} = 1$.

The action (\ref{IIBStringaction}) makes manifest an invariance under the SL(2,${\mathbb R}$) symmetry \cite{SWIIB}
\begin{eqnarray}
\tau \rightarrow \frac{a\tau +b}{c\tau +d},  \ \ \ F^i_3 \rightarrow \Lambda^i_j F^j_3, \cr
\tilde F_5 \rightarrow \tilde F_5,  \ \ \ g_E \rightarrow g_E,
\end{eqnarray}
for $\Lambda^i_j = \left[{d \atop b} {c \atop a}  \right]$ with $\det \Lambda = 1$.  The transformation with $\Lambda^i_j = \left[{0 \atop -1} {1 \atop 0}  \right]$ is of particular interest and is known as S-duality.  For solutions with $A_0 = 0$, it acts as $\phi \rightarrow - \phi$ and interchanges $F_3$ and $H_3$.  As a result, it maps D1-branes to fundamental strings and maps D5-branes to NS5-branes (and vice versa).  On the other hand, it leaves the D3-brane and all purely gravitatonal branes invariant.

Finally, we turn to T-duality, which is a symmetry that relates the full type IIA and IIB string theories. \index{T-duality} At the level of supergravity, this is a symmetry
that maps solutions of type IIA theory with a Killing field into
solutions of type IIB theory with a Killing field.  It appears
that T-duality is an exact symmetry of the underlying string/M-theory even with no Killing field, but that it then maps a nearly classical spacetime
into a complicated highly quantum mechanical state.

It is useful to first write down the explicit action of T-duality on
the metric and Neveu-Schwarz fields.  Let us introduce a coordinate $z$
such that translations in $z$ are a Killing symmetry.
Let $x^\alpha$
be any other collection of coordinates which makes $(z,x^\alpha)$ a
coordinate patch.  Again writing the anti-symmetric Neveu-Schwarz
field as $B_2$ instead of $A_2$, if the original solution is $(g,B)$
then the transformed solution $(\tilde g, \tilde B)$ is given \cite{Bus} by
\begin{eqnarray}
\label{Tdual}
\tilde g_{zz} &=& 1/g_{zz}, \ \ \ \tilde g_{z\alpha} = B_{z \alpha}/g_{zz}, \cr
\tilde g_{\alpha \beta } &=& g_{\alpha \beta} - ( g_{z\alpha}
g_{z\beta} - B_{z\alpha}B_{z\alpha})/g_{zz},  \ \ \
\tilde B_{z\alpha} = g_{z \alpha}/g_{zz}, \cr
\tilde B_{\alpha \beta} &=& B_{\alpha \beta} -  (g_{z\alpha}B_{\beta z}
- g_{z\beta}B_{\alpha z})/g_{zz}, \ \  \tilde{\phi} = \phi + \log g_{zz}.
\end{eqnarray}
Note that T-duality essentially interchanges the $g_{z \alpha}$
part of the metric with the $B_{z\alpha}$ part of the gauge field\footnote{In fact, in the case where the $z$-direction is compactified into an $S^1$, (\ref{Tdual}) can be described geometrically  by noting that the  Kaluza-Klein reduction to 8+1 dimensions has two 1-form gauge fields, one from the 10-dimensional metric and one from the 10-dimensional $B_2$. In the Neveu-Schwarz sector, T-duality simply interchanges the two circle-bundles associated with these gauge fields.}.    Now, in
the asymptotically flat context, the $g_{z0}$ component of the metric
is associated with momentum in the $z$-direction while the $B_{z0}$
component of the gauge field is associated with electrically charged
strings that extend in the $z$-direction.  Thus one finds \cite{HGS} that
T-duality interchanges momentum and charge, and in particular maps the (one-smeared)
Aichelburg-Sexl solutions to fundamental string solutions (which carry the electric charge to which $B$ couples).

Strictly speaking, the T-duality of string theory
requires a Killing field with compact
($S^1$) orbits, through (\ref{Tdual}) maps solutions to solutions even if the orbits are non-compact.  The original spacetime should be asymptotically Kaluza-Klein
and, if the original $z$ coordinate is identified such that the length
of the $S^1$ at infinity is $L$, then the $z$ coordinate of the
transformed spacetime should be identified such that the length of the
$S^1$ at infinity is $\frac{4 \pi^2 l_s^2}{L}$.  The point is that,
if the orbits of the Killing field are compact, then
quantum mechanics implies that the momentum component around the compact
direction is quantized.   Proper normalization then guarantees that
T-duality  takes a solution with one quantum of momentum
to a solution containing a single fundamental string.

The effect of T-duality on the Ramond-Ramond fields is as follows:
\begin{eqnarray}
\label{RRT}
\tilde F_{n,\alpha_1 ...\alpha_n} = (const) F_{n+1, z \alpha_1 ...\alpha_n},
\cr
\tilde F_{n,z \alpha_1 ...\alpha_{n-1}} = (const) F_{n-1, \alpha_1
...\alpha_{n-1}}.
\end{eqnarray}
Thus, if one takes a D$p$-brane and T-dualizes
in some direction along the brane, one obtains
a D$(p-1)$-brane solution which is smeared along the T-duality direction.
In string theory, D-brane charge is quantized.
The normalization constants in (\ref{RRT}) are chosen so that the transformed solution
has one unit of D$(p-1)$-brane charge when the original solution
has one unit of D$p$-brane charge.
Similarly, if one smears a unit charge
D$p$-brane solution in a transverse direction
keeping the total charge equal to one quantum,
the T-dual solution is a unit charge D$(p+1)$-brane.
See \cite{Joe,CVJ,BBS,DineBook,Kbook} for a discussion of D-brane tensions, charge
quantization, etc.  In any given direction, one may check that the T-duality transformation squares to the identity.

\section{Some remarks on string perturbation theory}
\label{pert}

The preceding sections we have discussed the supergravity aspects
of various string-theoretic branes, including D-branes. However,
the real power of D-branes, and thus their importance, stems
from the fact that there is a renormalizable (in fact, order by order finite)
quantum perturbation theory to complement the classical supergravity
description.  This perturbation theory
describes both the internal dynamics of D-branes and their interactions with
the supergravity fields.     In particular, it is the key to the famous
counting of states of BPS and near-BPS black holes.
Thus, although we will not discuss the details,
it is worthwhile to say a few words here about this perturbation
theory.   We hope this gives a useful complement to standard presentations
which concentrate more on the perturbation theory details.

\subsection{Background field expansions and perturbative string theory}
\label{bfe}

A useful framework from which to view string perturbation theory
is that of the background field expansion (see e.g. \cite{Bryce}).
Let us first review this idea in the context of standard quantum
field theory.  For definiteness, the reader may choose to focus on a
familiar low dimensional
interacting scalar field theory or even quantum mechanics.
We will use $\phi$ to denote
the scalar field or, more generally, as a schematic notation for
the collection of all relevant fields.

Let us begin by supposing that there is some complete quantum
theory of this field, consisting of a set of field operators
$\hat \phi(x)$ and an associated set of composite operators acting on
a Hilbert space.  Exact calculations for interacting quantum
field theories are seldom possible, and one must resort to various
approximation schemes and expansions in small parameters in order
to obtain results.  For situations where the field is nearly in
its vacuum state, standard perturbation theory (see e.g. \cite{ItzZub})
can be a useful technique.

However, this is not the only case
of interest.   For example, it may be that a laboratory device (or a
star, black hole, or astrophysical event) produces a large, essentially
classical, disturbance in the field $\phi$ and that one wishes to
study small quantum effects in the resulting behavior.  It is in
such a regime that background field methods are useful.
One first  considers the solution $\phi_0$ to the classical
field equations that would describe the situation if $\hbar$ were set
to zero.  One then rewrites the theory in terms of the field
$\widehat{\delta_0 \phi}(x) =
\hat \phi (x) - \phi_0(x)$.  Assuming that there is in fact a set
of semi-classical states in which the expectation value of
$\hat \phi(x)$ is close to $\phi_0(x)$ and in which the fluctuations
are `small,' it makes sense to attempt a perturbative treatment in
terms of the field $\widehat{\delta_0 \phi}$.

This is the
basic idea behind the background field expansion.  \index{background field expansion} However, there is
an additional subtlety.  Although
one expects any differences to vanish
as $\hbar \rightarrow 0$, there need not be any
state in which the expectation value of $\hat \phi(x)$ is exactly
$\phi_0(x)$.  In a perturbative framework, one assumes
that the difference between the actual expectation value and the
classical solution $\phi_0$ can be expanded in powers of $\hbar$
and simply solves for it at each order of
perturbation theory.  It is useful to take the expectation value
calculated at order $n$, which we write as $\overline \phi_n$, to
be an effective `classical field' and to work at order $n$ with the
perturbation $\widehat{\delta_n \phi} = \phi(x) - \overline \phi_n.$

Within the range of validity of this perturbation theory,
one can (see \cite{Bryce})
expand about a general classical solution $\phi_0$ and
obtain, at order $n$ in perturbation theory, an `effective action'
for the `effective classical background field' $\overline \phi$.
The variations of the effective action with respect to the effective
field yield the classical equations of motion for $\phi$ corrected
by terms of up to order $n$ in $\hbar$ such that the solutions
of these equations yield the expectation value of $\hat \phi(x)$
(to order $n$) in a semi-classical state.

One could also attempt to follow the same general framework
but to expand around some arbitrary field $\phi_0$ which is not
a solution to the classical equations of motion.  In this case,
the field $\widehat {\delta_0 \phi}(x)$ is not small and the perturbation
theory will not contain anything like a stable vacuum.  Because
the variation of the action does not vanish at the chosen
background, the action contains a term linear in
$\widehat {\delta_0 \phi}(x)$ which acts as a source.
This typically leads to
various infrared divergences in the perturbation theory since, when integrated
over all time, this source will produce an infinite number of particles.
Thus if, for some
reason, someone had handed us not the full classical dynamics of the field
but only the equations of the perturbation
theory around an arbitrary background, the classical solutions of the
theory would still be recognizable.

String perturbation theory
is in fact a version of background field theory in which the
`strings' correspond to excitations of the field $\widehat {
\delta_0 \phi} (x)$.  However, the logical order of the background
field framework is reversed or, perhaps more accurately, turned inside
out.  Instead of starting with a classical theory, quantizing,  and
performing the background field expansion, one instead postulates
the perturbative expansion\footnote{More
accurately, the S-matrix corresponding to such an expansion.}
about any background field
and then reconstructs the `classical' dynamics of the background field
in the manner discussed above
from the condition that the perturbation theory is well-defined.

This seemingly odd logical structure makes more sense when one
recalls that string theory is not, at present, a complete theory
based on any particular set of fundamental principles or axioms.
Rather, it is really an accidentally discovered set
of self-consistent mathematical phenomena
related to quantum gravity, the unification of forces, and so on.
The way that string perturbation theory arose historically was
through interest in QCD and possible `strings' of gauge field flux
that would connect quarks in hadrons.  While studying such strings, it
was discovered that they defined a perturbation theory which
was finite order by order and which contained a spin two particle
which could be interpreted as a graviton.  Since finding a
perturbative treatment of quantum gravity, or even constructing a new
theory of gravity which could be treated perturbatively, had
been a question of interest for some time, string theory
presented a solution to this technical problem:
Simply take this accidentally discovered
perturbation theory and use it to construct an associated theory of
quantum (and classical) gravity.  In the case of string theory,
the postulated perturbation theory was used to construct not
only the classical dynamics of the various fields, but also
to deduce the classical field content itself.  The rest, as they say,
is history.

Our story of supergravity discussed in the previous sections is relevant
here because the dynamics of string theory reduces in a certain
limit to classical supergravity.
A few fine points are worth mentioning briefly.  The first
is that, when viewed as a background field theory of the sort just discussed, classical string theory actually contains not only the
fields of classical supergravity, but an infinite tower of massive
fields as well.  The masses of these {\it classical} fields are, however,
on the order of $l_s^{-1}$ (and therefore considered to be large).  Thus,
one expects there to be a large
regime in which these fields are not independently excited.  Instead, the
heavy (massive) fields
are `locked' to the values of the massless fields.  At the extreme
end of this regime, the massive fields are completely irrelevant.
However, as one pushes toward the boundaries of this regime,
the massive fields may still have some effect on the dynamics.
If one solves the classical equations of motion for the heavy fields,
one finds that they are disturbed slightly by the massless fields and then
in turn provide small sources for the massless fields.  This analysis,
known in the lingo of path integrals
as `integrating out' the massive fields,
leads to additional effective interactions between the
massless fields.  Such interactions are
non-local on a scale set by the masses of the heavy fields; i.e., on the
scale of the string length.   When expanded in a power series, they
lead to a series of higher derivative terms in the action suppressed
by powers of the string scale.  These are the so-called
$\alpha'$-corrections, where $\alpha' \propto l_s^2$.

In this way the string scale explicitly appears in the dynamics of
classical string theory.  \index{string length} Now, it is true that in the `real world' the string
length is likely to be within a few orders of magnitude of
the Planck scale.  In principle, however,
the two scales are completely independent and should not be confused.
The string scale controls the
corrections to classical supergravity caused by the tower of massive
fields and the (9+1) Planck scale is the true quantum scale.
Their ratio defines the string coupling $g_s$. \index{string coupling}
The regime in which string perturbation theory is useful is $g_s \ll 1$,
in which the string length is much greater than the 9+1 Planck length.

\subsection{Strings and D-branes}
\label{SandD}

In order to describe how D-branes fit into this picture, we should say just
a few more words about the relation of strings to supergravity.
As mentioned above, strings provide
rules for constructing the perturbation theory about a given 9+1 supergravity
background.   Roughly speaking, one replaces
the Feynman diagrams (related to particles) of familiar perturbation theory
with a new sort of diagram related to strings.  For details, the
reader should consult \cite{GSW,Joe,CVJ,BBS,DineBook,Kbook}.
For most of our purposes below, it will
suffice to think about the strings as classical objects.

One can conceive of two basic types of strings.
The first are the so-called closed strings, \index{closed strings} which
at any moment of time have the topology $S^1$ and
resemble a classical rubber band.  It turns out that the closed
strings define a consistent perturbation theory in and of themselves, and
that it is this case that leads to the type II supergravities on which
we have focused.  Another version of the closed string leads
to heterotic supergravity, which has half as much supersymmetry
as the type II theories.

One might also consider so-called open strings which, \index{open strings}
at any instant of time, have the topology of an interval.  In order
for the dynamics of such strings to be well-defined, one must
specify boundary conditions at the ends.    A natural
choice is to impose Neumann boundary conditions to describe
free ends.  Such strings are quite similar to classical rubber bands
that have been cut open.  It turns out
that this type of string does not yield a consistent perturbation theory by
itself, as two open strings can join together to produce a closed
string.  When open and closed strings are taken together, a consistent
perturbation theory does result.  This theory is associated with
type I supergravity, having half as much supersymmetry as the  type II
theories.

The other type of boundary condition that one can impose at the end
of a string is the Dirichlet boundary condition, requiring the end
of the string to remain fixed at some point in space.  One can also
consider a mixture of Dirichlet and Neumann boundary conditions,
insisting that the end of the string remain attached to some submanifold
of spacetime, but otherwise leaving it free to roam around the
surface.  Surfaces associated with such Dirichlet boundary conditions
are known as Dirichlet submanifolds, or D-branes.
Again, for a consistent perturbation theory,  one must
consider closed strings in addition to these open strings.
Since we have singled out this submanifold as a special
place in the spacetime, this perturbation theory should not
describe an expansion about empty space.  However,
there remains the possibility
that it can describe an expansion about a background in which
certain sub-manifolds are picked out as special; i.e., near
a background which includes certain brane-like features.
Recall that, as a background field expansion, this perturbation
theory should tell us about all of the dynamics of the background,
including any dynamics of the branes.

To make a long story short, it turns out that the Dirichlet submanifolds
are sources of the Ramond-Ramond gauge fields and of the gravitational
field.  That is, they carry both stress-energy and Ramond-Ramond
charge.  Thus, one might expect that they have something to
do with the branes discussed earlier that carry Ramond-Ramond charge.
In fact, this D-brane perturbation theory is supposed to give the
expansion about a background that includes such a
charged gravitating R-R brane in the regime of asymptotically small string
coupling $g_s$, which controls the strength of all interactions.
The perturbation theory describes both the dynamics
of the bulk fields (roughly speaking,
through the closed strings) and of the brane
itself (roughly speaking, through the open strings).
The two parts are coupled and interact.

Let us briefly comment on how this picture meshes with the supergravity point of view.  To do so, we must consider the strength of the above interaction.  This is determined by the source the brane provides for the various supergravity fields.  But a simple brane has only one charge, which is necessarily equal to its mass per unit volume if the brane is BPS.  As a result, the strength of the source is governed by $GT$ where $16\pi G=2\kappa^2_{10} =  (2\pi)^7 g_s^2 l_s^8$
%is the 10-dimensional Newton's constant
and $T$ is the brane tension.

Now, in string theory,
as in any theory with both
electric and magnetic charges, the charge (and thus $T$ for BPS branes) is quantized in integer
multiples of some fundamental unit.
It turns out that the charge of any R-R brane (with $n$
units of charge) is proportional to $n/g_s$.  Thus, $GT \sim n g_s$
goes to zero at weak coupling.  As a result, the supergravity
fields go over to flat empty space in this limit and the field generated by a fixed number $n$ of such branes is indeed perturbative at small $g_s$.  On the other hand, since the mass per unit
volume of the D-brane is diverging, any internal dynamics associated
with motion of the D-brane
is frozen out in this limit.  Thus the picture from supergravity agrees with the string perturbation theory described above:  up to perturbations, it consists
of flat empty space with a preferred submanifold in spacetime occupied
by a largely non-dynamical brane.  This suggests that the D-branes
of perturbation theory should be identified with the
Ramond-Ramond branes of supergravity.  Additional evidence for this
picture comes from the great success of D-brane perturbation theory
in reproducing the entropy of black
holes \cite{SV,CM,HS2,BMPV,MS,JKM,BLMPSV,KT,JM,HM1,HM2}, Hawking radiation
\cite{DMW,DM1,DM2}, and
the so-called grey-body factors \cite{MSII}
associated with the Ramond-Ramond branes.

In contrast, the Neveu-Schwarz branes do not admit simple descriptions as backgrounds for string perturbation theory. For the fundamental string, this is because  (with $n$ units of charge)
the tension is proportional to $n$ and does not depend on the string
coupling.  They therefore remain fully dynamical at small $g_s$ (while the spacetime solution again becomes just Minkowski space).  For the NS5-brane (with $n$ units of charge) the
tension is proportional to $n/g_s^2$.  Thus, $GT \sim n$
and the supergravity fields remain unchanged as we take
$g_s \rightarrow 0$; the fields cannot be described as small perturbations about Minkowski space.

\subsection{A few words on black hole entropy}\index{entropy}
\label{entropy}

This is not the place for an in-depth discussion of just how D-brane
perturbation theory can be used to reproduce the properties of
supergravity solutions.  Such treatments can be found in \cite{JMPHD}
and in \cite{Joe,CVJ,BBS,Kbook}.  They involve the fact that the open strings
associated with D-branes describe, in the low energy limit, a certain
non-abelian Yang-Mills theory.  The low energy limit of that theory
can then be analyzed and used to study the low energy limit of the
brane dynamics.  Since BPS branes have the minimal possible energy
for their charge, this means that BPS and nearly BPS branes can
be addressed by such techniques.

We will, however, close by giving some parts of the entropy calculation for
a particular case.   As has already been mentioned, the solution
(\ref{3M2}) with three mutually orthogonal sets of M2-branes is
the simplest BPS black brane solution with non-zero entropy.  Let us compactify
a circle along one of the M2-branes (say, the one associated with the
$z_\parallel$ coordinates) and Kaluza-Klein reduce to 9+1 type IIA
supergravity.  Then, as we have seen, the $z$-type M2-branes (which are
wrapped around this circle)
become fundamental strings in the IIA description while the
$x$- and $y$-type M2-branes (which do not wrap around the compact circle)
become D2-branes.  It turns out that a simple description
of the microscopic perturbative states can be obtained by T-dualizing
this solution to the IIB theory along the direction in which the
fundamental strings point.  This turns the fundamental strings into
momentum and the two sets of D2-branes into D3-branes.

Let us now
T-dualize twice more in, say, the two $y_\parallel$ directions.
This again yields a solution of IIB theory.  The momentum remains
momentum in the same direction, but one of the sets of D3-branes
has become a set of D1-branes and the other has become a set of D5-branes.
The D1-branes (D-strings) are stretched in the
same direction that the momentum is flowing, and this all happens
in one of the directions along the D5-branes.  These T-dualities
do not change the integer charges $Q_x,Q_y,Q_z$ associated with the various
types of branes: $Q_x$ is now the number of D5-branes, $Q_y$ the number
of D-strings, and $Q_z$ the number of momentum quanta.  One can check
that these T-dualities do not change the Bekenstein-Hawking entropy and,
as supposed symmetries of the underlying string theory, they cannot
change the number of microstates.

The case of a single D5-brane is particularly simple
to discuss.  It turns out that
the low energy dynamics
reduces to what is effectively just a collection of D-strings\footnote{Or,
even better, to a single D-string wrapped $Q_y$ times
around the direction in which the momentum flows. See, e.g.,
\cite{JMPHD}.} which
are stuck to the D5-brane but free to oscillate within it.  The momentum
in the solution is just the momentum carried by these oscillations, and
the energy of the solution is a linear sum of contributions from the
D5-brane rest energy, the D-string rest energy, and the momentum.
For a supersymmetric solution, all of the oscillations must move
in the same direction along the D-string and so are described by, say,
right-moving fields on a 1+1 dimensional spacetime.
Oscillations of D-strings propagate at the speed of light and so the
associated energy-momentum vector is null.

Thus, for
each string, one has $4$ massless $1+1$ rightmoving scalar fields corresponding
to the four internal directions of the fivebrane.  Supersymmetry
implies that there
are also four massless $1+1$ rightmoving
fermionic fields for each D-string.  A fermion
acts roughly like half of a boson, so we may think of this as $6Q_y$ massless
right-moving scalars
on $S^1 \times R$ (the worldvolume of a one-brane).
A standard formula tells us that, given $n$ massless rightmoving scalars
with $Q_z$ units of momentum,
the entropy at large $Q_z$
is $S = 2\pi \sqrt{Q_z n/6}$.  Thus we have
$S = 2\pi \sqrt{Q_yQ_z}$, in agreement
with the Bekenstein-Hawking entropy $S = A/4G_{11}$ (see
eq. (\ref{BHE}) with $Q_x =1$) for the associated black hole.

This gives an idea of the way in which D-brane perturbation theory provides
a microscopic accounting of the entropy of this BPS black hole.
The other BPS and near-BPS cases are similar in many respects.
It is quite satisfying to arrive at exactly the Bekenstein-Hawking
entropy formula without having to adjust any free parameters.
However, one is certainly struck by the qualitative differences
between the regime in which we are used to thinking about black holes and
the regime in which the string calculation is performed.  We usually
consider black holes with large smooth horizons.  In contrast, the perturbative
calculation is done in the asymptotic regime of small $g_s$,
where spacetime is flat and the horizon has degenerated to zero size.
The belief is that supersymmetry guarantees the entropy of the quantum
system to be independent of $g_s$, as it does for other non-gravitational
systems\footnote{As supporting evidence, recall that the Bekenstein-Hawking
entropy of our BPS black hole does not depend on $g_s$ when written
in terms of the integer charges (\ref{BHE}).}.
However, there is much room for speculation and investigation
in trying to match these pictures more closely and in understanding just
what form these states take in the black hole regime of finite $g_s$.

\vskip 1cm

{\centerline {\bf Acknowledgements}}

The author wishes to thank Gary Horowitz, Clifford Johnson, Rob Myers,
Amanda Peet, and especially Joe
Polchinski for innumerable discussions on string theory and stringy black holes, as well as  Tomas Andrade, Tony Zee, and especially Gary Horowitz for spotting numerous typos and errors in earlier versions of this chapter.
This work was supported in part by the US National Science Foundation under grant PHY08-55415 and by funds from the University of California.

%\endinput

\begin{thebibliography}{99}
\bibitem{YR} D. Youm,  ``Black Holes and solitons in string
theory,'' {\it Phys. Rept.} {\bf 316} 1 (1999) hep-th/9710046.

\bibitem{45rev}
  K.~-i.~Maeda, M.~Nozawa,
  ``Black hole solutions in string theory,''
  [arXiv:1104.1849 [hep-th]].

\bibitem{ResLet}
  D.~Marolf,
  ``Resource letter: The Nature and status of string theory,''
  Am.\ J.\ Phys.\  {\bf 72}, 730-741 (2004).
  [hep-th/0311044].

\bibitem{GSW} M. Green, J. Schwarz, and E. Witten, {\it Superstring
theory}, (Cambridge U. Press, NY, 1987).

\bibitem{Joe} J. Polchinski, {\it String Theory} (Cambridge U. Press,
Cambridge, 1998).



\bibitem{CVJ}
 C.~V.~Johnson,
  ``D-branes,''
  Cambridge, USA: Univ. Pr. (2003) 548 p.


\bibitem{BBS}
K.~Becker, M.~Becker, J.~H.~Schwarz,
  ``String theory and M-theory: A modern introduction,''
  Cambridge, UK: Cambridge Univ. Pr. (2007) 739 p.

\bibitem{DineBook}
  M.~Dine,
  ``Supersymmetry and string theory: Beyond the standard model,''
  Cambridge, UK: Cambridge Univ. Pr. (2007) 515 p.

\bibitem{Kbook}
  E.~Kiritsis,
  ``String theory in a nutshell,''
  Princeton, USA: Univ. Pr. (2007) 588 p.




\bibitem{CJS} E. Cremmer, B. Julia, and J. Scherk,  ``Supergravity
Theory in Eleven-Dimensions,'' {\it Phys. Lett.}
{\bf B76} 409 (1978).


\bibitem{mult} P. C. Aichelburg and F. Embacher, ``Exact
Superpartners of N=2 supergravity solitons,'' {\it Phys.
Rev. D} {\bf 34} 3006 (1986); M. J. Duff, J. T. Liu, and J. Rahmfeld,
``Dipole Moments of Black Holes and String States,'' {\it Nucl. Phys.}
{\bf B494} 161 (1996) hep-th/9612015;  M. J. Duff, J. T. Liu, and
J. Rahmfeld, ``g=1 for Dirichlet 0-branes,'' {\it Nucl.
Phys.} {\bf B524} 129 (1998)  hep-th/9801072;
V. Balasubramanian, D. Kastor, and J. Traschen,
``The spin of the M2-brane and spin-spin interactions via
probe techniques,'' {\it Phys. Rev. D} {\bf 59} 984007 (1999).

\bibitem{Town} Townsend, P.K. ``Brane Surgery,"
{\it Nucl. Phys. Proc. Suppl.} {\bf 58} 163 (1997).

\bibitem{3Q}
  D.~Marolf,
  ``Chern-Simons terms and the three notions of charge,''
in {\it Quantization, Gauge
Theory, and Strings -- Proceedings of the International
Conference dedicated to the memory of Professor Efim Fradkin}, editors
A. Semikhatov, M. Vasiliev, and V. Zaiken (Scientific World, Moscow, 2001);
[hep-th/0006117].

\bibitem{GH} G. W. Gibbons and C. M. Hull, ``A Bogomolny Bound
for General Relativity and Solitons in N=2 Supergravity,''
{\it Phys. Lett.} {\bf 109B} 190 (1982).

\bibitem{Witten} E. Witten, ``A simple proof of the positive energy
theorem,'' {\it Comm. Math. Phys.} {\bf 80} 381 (1981).

\bibitem{AGT}
 J.~A.~de Azcarraga, J.~P.~Gauntlett, J.~M.~Izquierdo, P.~K.~Townsend,
  ``Topological Extensions of the Supersymmetry Algebra for Extended Objects,''
  Phys.\ Rev.\ Lett.\  {\bf 63}, 2443 (1989).

\bibitem{GHT}  G. W. Gibbons, G. T. Horowitz, and P. K. Townsend,
``Higher-dimensional resolution of dilatonic black-hole singularities,''
{\it Class. Quant. Grav.} {\bf 12} 297 (1995).


\bibitem{alg} E. Witten and  D. Olive,
``Supersymmetry algebras that include topological charges,''
{\it Phys. Lett.} {\bf 78B} (1978) 97;
S. Ferrara, C. A. Savoy, and B. Zumino, ``General Massive
Multiplets in Extended Supersymmetry,'' {\it Phys. Lett.} {\bf 100B}
(1981) 393.


\bibitem{MP}  S. Majumdar, {\it Phys. Rev.} {\bf 72} (1947) 930;
A. Papapetrou, {\it Proc. Roy. Irish Acad.} {\bf A51}
(1947) 191.

\bibitem{HHMP}
 J.~B.~Hartle, S.~W.~Hawking,
  ``Solutions of the Einstein-Maxwell equations with many black holes,''
  Commun.\ Math.\ Phys.\  {\bf 26}, 87-101 (1972).

\bibitem{Welch}
D.~L.~Welch,
  ``On the smoothness of the horizons of multi - black hole solutions,''
  Phys.\ Rev.\  D {\bf 52}, 985 (1995)
  [arXiv:hep-th/9502146].
  %%CITATION = PHRVA,D52,985;%%

\bibitem{GCHR}
  G.~N.~Candlish, H.~S.~Reall,
  ``On the smoothness of static multi-black hole solutions of higher-dimensional Einstein-Maxwell theory,''
  Class.\ Quant.\ Grav.\  {\bf 24}, 6025-6040 (2007).
  [arXiv:0707.4420 [gr-qc]].




\bibitem{PI}  E.~Poisson and W.~Israel,
  ``Internal structure of black holes,''
  Phys.\ Rev.\  D {\bf 41}, 1796 (1990).
  %%CITATION = PHRVA,D41,1796;%%


\bibitem{Dafermos}
  M.~Dafermos,
  ``The interior of charged black holes and the problem of uniqueness in
  general relativity,'' Communications on Pure and Applied Mathematics, {\bf 58}, 445-504, (2005)  [arXiv:gr-qc/0307013].

\bibitem{BradySmith}
P.~R.~Brady and J.~D.~Smith,
  ``Black Hole Singularities: A Numerical Approach,''
  Phys.\ Rev.\ Lett.\  {\bf 75}, 1256 (1995)
  [arXiv:gr-qc/9506067].
  %%CITATION = PRLTA,75,1256;%%

\bibitem{extremes}
  D.~Marolf,
  ``The dangers of extremes,''
  Gen.\ Rel.\ Grav.\  {\bf 42}, 2337-2343 (2010).
  [arXiv:1005.2999 [gr-qc]]; D. Marolf and A. Ori, to appear.

\bibitem{AS} P. C. Aichelburg, R. U. Sexl, ``On the Gravitational
Field of a Massless Particle,'' {\it Gen. Rel. Grav.} {\bf 2}
303 (1971).

\bibitem{M2sol} E. Bergshoeff, E. Sezgin, and P.K. Townsend,
``Supermembranes and eleven-dimensional supergravity,'' {\it
Phys. Lett.} {\bf B189} 75 (1987).

\bibitem{Rafael} R. Sorkin, ``Kaluza-Klein monopole,''
{\it Phys. Rev. Lett.} {\bf 51} 87 (1983).

\bibitem{GP} D. Gross and M. Perry, ``Magnetic Monopoles in Kaluza-Klein
Theories,'' {\it Nucl. Phys.} {\bf B226} 29 (1983).



\bibitem{HawkingEllis} S. Hawking and G. F. R. Ellis, {\it
The large-scale structure of space-time} (Cambridge University Press, 1973).

\bibitem{NUT} E. T. Newman, L. Tamburino, and T. Unti, {\it J. Math.
Phys.} {\bf 4} (1963) 915.

\bibitem{BTZ} M. Ba\~nados, C. Teitelboim, and J. Zanelli, ``The
black hole in three-dimensional space-time,'' {\it Phys. Rev. Lett.}
{\bf 69} 1849 (1992), hep-th/9204099;
M. Ba\~nados, M. Henneaux, C. Teitelboim, and J. Zanelli, ``Geometry
of the (2+1) black hole,'' {\it Phys. Rev. D}
{\bf 48} 1506 (1993), gr-qc/9302012.

\bibitem{Tharm}
  A.~A.~Tseytlin,
  ``Harmonic superpositions of M-branes,''
  Nucl.\ Phys.\  B {\bf 475}, 149 (1996)
  [arXiv:hep-th/9604035].
  %%CITATION = NUPHA,B475,149;%%

\bibitem{GKT}
  J.~P.~Gauntlett, D.~A.~Kastor, J.~H.~Traschen,
  ``Overlapping branes in M theory,''
  Nucl.\ Phys.\  {\bf B478}, 544-560 (1996).
  [hep-th/9604179].

\bibitem{angles}  K. Behrndt and M. Cvetic, ``BPS Saturated Bound
States of Tilted p-branes in type II string theory,''
{\it Phys. Rev. D} {\bf 56} (1997) 1188-1193, hep-th/9702205;
J.C. Breckenridge, G. Michaud, and R.C. Myers, ``New angles on D-branes,''
{\it Phys. Rev.} {\bf D56} (1997) 5172-5178,
hep-th/9703041; V. Balasubramanian, F. Larsen, and R.G. Leigh, ``Branes
at angles and black holes,''
{\it Phys. Rev.} {\bf D57} (1998) 3509-3528,
hep-th/9704143.

\bibitem{SS} S. Surya and D. Marolf, ``Localized Branes and Black Holes,''
Phys.Rev. D58 (1998) 124013, hep-th/9805121.

\bibitem{AP}   D.~Marolf, A.~W.~Peet,
  ``Brane baldness versus superselection sectors,''
  Phys.\ Rev.\  {\bf D60}, 105007 (1999).
  [hep-th/9903213].


\bibitem{pertdiv} A.~Gomberoff, D.~Kastor, D.~Marolf, J.~H.~Traschen,
  ``Fully localized brane intersections - the plot thickens,''
  Phys.\ Rev.\  {\bf D61}, 024012 (2000).
  [hep-th/9905094].



\bibitem{CHI}
  S.~A.~Cherkis, A.~Hashimoto,
  ``Supergravity solution of intersecting branes and AdS/CFT with flavor,''
  JHEP {\bf 0211}, 036 (2002).
  [hep-th/0210105].

\bibitem{DEG1}
  E.~D'Hoker, J.~Estes, M.~Gutperle,
  ``Exact half-BPS Type IIB interface solutions. II. Flux solutions and multi-Janus,''
  JHEP {\bf 0706}, 022 (2007).
  [arXiv:0705.0024 [hep-th]].

\bibitem{DEG2}
  E.~D'Hoker, J.~Estes, M.~Gutperle,
  ``Exact half-BPS Type IIB interface solutions. I. Local solution and supersymmetric Janus,''
  JHEP {\bf 0706}, 021 (2007).
  [arXiv:0705.0022 [hep-th]].

\bibitem{CWIIA}
 I.~C.~G.~Campbell, P.~C.~West,
  ``N=2 D=10 Nonchiral Supergravity and Its Spontaneous Compactification,''
  Nucl.\ Phys.\  {\bf B243}, 112 (1984).

\bibitem{HNIIA}   M.~Huq, M.~A.~Namazie,
  ``Kaluza-klein Supergravity In Ten-dimensions,''
  Class.\ Quant.\ Grav.\  {\bf 2}, 293 (1985).

\bibitem{GPIIA}
  F.~Giani, M.~Pernici,
  ``N=2 Supergravity In Ten-dimensions,''
  Phys.\ Rev.\  {\bf D30}, 325-333 (1984).

\bibitem{attract1}
S. Ferrara, R. Kallosh and A. Strominger, "N=2 extremal black holes," Phys. Rev. D
52, 5412 (1995) arXiv:hep-th/9508072.

\bibitem{attract2}
S. Ferrara and R. Kallosh, "Supersymmetry and Attractors," Phys. Rev. D 54 (1996)
1514, arXiv:hep-th/9602136;
S. Ferrara, G. W. Gibbons and R. Kallosh, "Black holes and critical points in moduli
space," Nucl. Phys. B 500, 75 (1997) [arXiv:hep-th/9702103];
S. Ferrara, K. Hayakawa and A. Marrani, "Lectures on Attractors and Black Holes,"
Fortsch. Phys. 56, 993 (2008) [arXiv:0805.2498 [hep-th]];
S. Bellucci, S. Ferrara, M. Gunaydin and A. Marrani, "SAM Lectures on Extremal
Black Holes in d=4 Extended Supergravity," arXiv:0905.3739 [hep-th];
T. Ortin, "Supersymmetric solutions of 4-dimensional supergravities," AIP Conf.
Proc. 1318, 175 (2010) [arXiv:1010.1383 [gr-qc]];   S.~Kachru, R.~Kallosh, M.~Shmakova,
  ``Generalized Attractor Points in Gauged Supergravity,''
  [arXiv:1104.2884 [hep-th]].

\bibitem{non-BPS} O. Bergman, M. R. Gaberdiel, ``A nonsupersymmetric
open string theory and S-duality,'' {\it Nucl. Phys.} {\bf B499}
193 (1997), hep-th/9701137;
O. Bergman, M. R. Gaberdiel, ``Stable non-BPS D-particles,''
{\it Phys. Lett.} {\bf B441} 133 (1998), hep-th/9806155;
O. Bergman, M. R. Gaberdiel, ``Non-BPS states in Heterotic type IIA
duality,'' {\it JHEP} {\bf 9903} 013 (1999).

\bibitem{non-BPSII} A. Sen, ``Stable Non-BPS states in string theory,''
{\it JHEP} {\bf 9806} 007 (1998), hep-th/9803194; A. Sen,
``Stable Non-BPS bound states of D-branes,'' {\it JHEP}
{\bf 9808} 010 (1998), hep-th/9805019; A. Sen and B.
Zweibach, ``Stable Non-BPS states in F
Theory,'' hep-th/9907164; M. R. Gaberdiel and A. Sen, ``Nonsupersymmetric
D-brane configurations with Bose-Fermi Degenerate
open string spectrum,'' hep-th/9908060.

\bibitem{HSbs}
  G.~T.~Horowitz, A.~Strominger,
  ``Black strings and P-branes,''
  Nucl.\ Phys.\  {\bf B360}, 197-209 (1991).

\bibitem{SWIIB} J.~H.~Schwarz, P.~C.~West,
  ``Symmetries and Transformations of Chiral N=2 D=10 Supergravity,''
  Phys.\ Lett.\  {\bf B126}, 301 (1983).

\bibitem{HWIIB}
 P.~S.~Howe, P.~C.~West,
  ``The Complete N=2, D=10 Supergravity,''
  Nucl.\ Phys.\  {\bf B238}, 181 (1984).

\bibitem{JSIIB}
  J.~H.~Schwarz,
  ``Covariant Field Equations of Chiral N=2 D=10 Supergravity,''
  Nucl.\ Phys.\  {\bf B226}, 269 (1983).

\bibitem{Bus} T. Buscher, ``Path Integral Derivation of Quantum Duality
in Nonlinear Sigma Models," {\it Phys. Lett.} {\bf B201} 466 (1988);
``A Symmetry of the String Background Field Equations," {\it Phys. Lett.}
{\bf B194} 59 (1987).

\bibitem{HGS} J. H. Horne, G. T. Horowitz, A. R. Steif,
``An Equivalence Between Momentum and Charge in String Theory,''
{\it Phys. Rev. Lett.} {\bf 68} (1992) 568-571; hep-th/9110065.


\bibitem{Bryce} B. DeWitt, {\it Relativity, Groups, and Topology II,
Proceedings of the 1983 Les Houches Summer School}, ed. by
B. DeWitt and R. Stora (Elsevier Science Publishers, Amsterdam, 1984).

\bibitem{ItzZub} C. Itzykson and J.-B. Zuber, {\it Quantum Field Theory}
(McGraw-Hill, New York, 1980); S. Weinberg, {\it The Quantum Theory of
Fields} (Cambridge University Press, New York, 1995).

\bibitem{SV} A. Strominger and C. Vafa, ``Microscopic
Origin of the Bekenstein-Hawking Entropy,'' {\it Phys. Lett. B}
{\bf 379} (1996) 99-104, hep-th/9601029.

\bibitem{CM}  C. Callan and J. Maldacena, ``D-brane
approach to black hole quantum mechanics,'' {\it Nucl Phys}
{\bf B472} (1996) 591-610, hep-th/9602043.

\bibitem{HS2} G. Horowitz and A. Strominger, ``Counting
states of near extremal black holes,'' {\it Phys. Rev. Lett.}
{\bf 77} (1996) 2368, hep-th/9602051.

\bibitem{BMPV} J. Breckenridge, R. Myers, A. Peet and C. Vafa, ``D-branes
and spinning black holes,''
{\it Phys. Lett. B} {\bf 391} (1997) 93-98,  hep-th/9602065.

\bibitem{MS}  J. Maldacena and A. Strominger, ``Statistical
Entropy of Four-dimensional extremal black holes,'' {\it Phys. Rev. Lett.}
{\bf 77} (1996) 428-429, hep-th/9603060.

\bibitem{JKM}  C. Johnson, R. Khuri, and R. Myers, ``Entropy of 4-D
extremal black holes,'' {\it Phys.
Lett. B} {\bf 378} (1996) 78-86, hep-th/9603061.
i
\bibitem{BLMPSV}  J. Breckenridge, D. Lowe, R. Myers, A.  Peet,
A. Strominger and C. Vafa, ``Microscopic entropy of near
extremal spinning black holes,'' {\it Phys Lett. B} {\bf 381}
(1996) 423-426, hep-th/9603078.

\bibitem{KT}
I.R. Klebanov, A.A. Tseytlin, ``Intersecting M-branes as four-dimensional
black holes,''
{\it Nucl. Phys.} {\bf B475} (1996) 179-192,
hep-th/9604166.

\bibitem{JM}
J. M. Maldacena, ``N=2 Extremal Black Holes and Intersecting Branes,''
{\it Phys. Lett.} {\bf B403} (1997) 20-22,
hep-th/9611163.

\bibitem{HM1} G.T. Horowitz and D. Marolf, ``Counting States of Black
Strings with Traveling waves,''
{\it Phys. Rev. D} {\bf 55} (1997) 835-845,
hep-th/9605224.

\bibitem{HM2} G. T. Horowitz and D. Marolf, ``Counting States of Black
Strings with Traveling waves II,''
{\it Phys. Rev. D} {\bf 55} (1997) 846-852,
hep-th/9606113.

\bibitem{DMW} A. Dhar, G. Mandal, and S. Wadia,
``Absorption vs. Decay of Black Holes in String Theory and
T symmetry,'' {\it Phys. Lett.}
{\bf B388} (1996) 51, hep-th/9605234.

\bibitem{DM1} S.R. Das and S.D. Mathur, ``Comparing Decay
Rates for Black Holes and D-branes,'' {\it Nucl. Phys.} {\bf
B478} (1996) 561-576,
hep-th/9606185.

\bibitem{DM2} S.R. Das and S.D. Mathur, ``Interactions Involving
D-branes,'' {\it  Nucl. Phys.} {\bf B482}
(1996) 153-172,
hep-th/9607149.


\bibitem{MSII} J. Maldacena and A. Strominger,
``Black Hole Greybody Factors and D-Brane Spectroscopy,'' {\it Phys. Rev.}
{\bf D55} (1997) 861-870, hep-th/9609026.


\bibitem{JMPHD} J. Maldacena, ``Black Holes in String Theory,''
Ph.D. thesis, hep-th/9607235.


\end{thebibliography}
\end{document}